\documentclass[fleqn,usenatbib]{rasti}
\usepackage{newtxtext,newtxmath}
\usepackage[T1]{fontenc}
\usepackage{natbib}
\DeclareRobustCommand{\VAN}[3]{#2}
\let\VANthebibliography\thebibliography
\def\thebibliography{\DeclareRobustCommand{\VAN}[3]{##3}\VANthebibliography}
\usepackage{graphicx}	
\usepackage{amsmath}	
\usepackage{enumitem}
\usepackage{hyperref}

\setenumerate{label*=\arabic*.}
\usepackage{caption}
\usepackage{cuted}
\usepackage{stfloats}

\usepackage[most]{tcolorbox}

\usepackage{minted}

\usepackage{xcolor}
\usepackage{setspace}
\usepackage{enumitem}  
\usepackage{etoolbox}  
\definecolor{darkblue}{RGB}{0,0,139}
\PassOptionsToPackage{colorlinks=true,urlcolor=darkblue}{hyperref}
\usepackage{hyperref}

\newcommand{\orchestralnestindent}{1.5em}  

\newcommand{\orchestralfontsize}{\small}  

\newcommand{\orchestralcodestyle}{vs}  

\newcommand{\orchestralitemsep}{0pt}           
\newcommand{\orchestralparsep}{0pt}            
\newcommand{\orchestraltopsep}{0pt}            
\newcommand{\orchestralpartopsep}{0pt}         

\definecolor{OrcUserColor}{RGB}{60, 60, 60}         
\definecolor{OrcAgentColor}{RGB}{120, 120, 255}     
\definecolor{OrcSystemColor}{RGB}{180, 0, 180}      
\definecolor{OrcToolColor}{RGB}{200, 100, 0}        

\definecolor{OrcPendingColor}{RGB}{128, 128, 128}   
\definecolor{OrcFailedColor}{RGB}{220, 20, 60}      

\definecolor{OrcCodeBg}{RGB}{36, 41, 51}            

\newtcolorbox{orchestralusermessage}{%
  enhanced,
  breakable,
  colback=white,
  colframe=OrcUserColor,
  boxrule=0.6pt,
  arc=1.5mm,
  width=\linewidth,
  top=8pt,
  bottom=8pt,
  left=10pt,
  right=10pt,
  fontupper=\orchestralfontsize,
  attach boxed title to top left={
    yshift=-6pt,
    xshift=3.5mm
  },
  boxed title style={
    colback=white,
    colframe=white,
    boxrule=0pt,
    left=0pt,
    right=0pt,
    top=3pt,
    bottom=0pt,
    sharp corners,
  },
  coltitle=OrcUserColor,
  fonttitle=\sffamily,
  title={\raisebox{0pt}[0pt][0pt]{User}},
}

\newtcolorbox{orchestralagentmessage}{%
  enhanced,
  breakable,
  colback=white,
  colframe=OrcAgentColor,
  boxrule=0.6pt,
  arc=1.5mm,
  width=\linewidth,
  top=8pt,
  bottom=8pt,
  left=10pt,
  right=10pt,
  fontupper=\orchestralfontsize,
  attach boxed title to top left={
    yshift=-6pt,
    xshift=3.5mm
  },
  boxed title style={
    colback=white,
    colframe=white,
    boxrule=0pt,
    left=0pt,
    right=0pt,
    top=3pt,
    bottom=0pt,
    sharp corners,
  },
  coltitle=OrcAgentColor,
  fonttitle=\sffamily,
  title={\raisebox{0pt}[0pt][0pt]{Agent}},
}

\newtcolorbox{orchestralsystemmessage}{%
  enhanced,
  breakable,
  colback=white,
  colframe=OrcSystemColor,
  boxrule=0.6pt,
  arc=1.5mm,
  width=\linewidth,
  top=8pt,
  bottom=8pt,
  left=10pt,
  right=10pt,
  fontupper=\orchestralfontsize,
  attach boxed title to top left={
    yshift=-6pt,
    xshift=3.5mm
  },
  boxed title style={
    colback=white,
    colframe=white,
    boxrule=0pt,
    left=0pt,
    right=0pt,
    top=3pt,
    bottom=0pt,
    sharp corners,
  },
  coltitle=OrcSystemColor,
  fonttitle=\sffamily,
  title={\raisebox{0pt}[0pt][0pt]{System}},
}

\newtcolorbox{orchestraltoolmessage}[1]{%
  enhanced,
  breakable,
  colback=white,
  colframe=OrcToolColor,
  boxrule=0.5pt,
  arc=1mm,
  width=\linewidth,
  left=\orchestralnestindent,
  right=1em,
  top=6pt,
  bottom=6pt,
  fontupper=\orchestralfontsize,
  attach boxed title to top left={
    yshift=-5pt,
    xshift=3.5mm
  },
  boxed title style={
    colback=white,
    colframe=white,
    boxrule=0pt,
    left=0pt,
    right=0pt,
    top=3pt,
    bottom=0pt,
    sharp corners,
  },
  coltitle=OrcToolColor,
  fonttitle=\sffamily\orchestralfontsize,
  title={\raisebox{0pt}[0pt][0pt]{#1}},
}

\newminted{python}{
  style=\orchestralcodestyle,
  fontsize=\small,
  breaklines,
  autogobble
}

\newminted{bash}{
  style=\orchestralcodestyle,
  fontsize=\small,
  breaklines,
  autogobble
}

\setlist[enumerate]{
  itemsep=\orchestralitemsep,
  parsep=\orchestralparsep,
  topsep=\orchestraltopsep,
  partopsep=\orchestralpartopsep
}

\setlist[itemize]{
  itemsep=\orchestralitemsep,
  parsep=\orchestralparsep,
  topsep=\orchestraltopsep,
  partopsep=\orchestralpartopsep
}

\setminted{
  style=vs,
  fontsize=\orchestralfontsize,
  frame=lines,
  bgcolor=gray!5,
  xleftmargin=0pt,
  xrightmargin=0pt,
  tabsize=4,
  showtabs=false
}

\AtBeginEnvironment{minted}{\vspace{0.3\baselineskip}}
\AtEndEnvironment{minted}{\vspace{0.3\baselineskip}}

\title[ASTER - Agentic Science Toolkit for Exoplanet Research]{\includegraphics[height=6em]{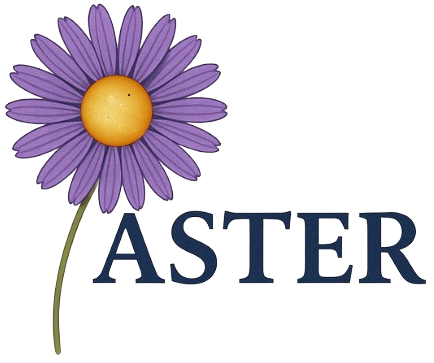} \raisebox{25pt}{ - Agentic Science Toolkit for Exoplanet Research}}

\author[E. Panek et al.]{Emilie Panek,$^{1}$\thanks{E-mail: epanek1@ua.edu}
Alexander Roman,$^{1}$
Gaurav Shukla,$^{1}$
Leonardo Pagliaro,$^{2}$
Katia Matcheva$^{1}$
and \newauthor Konstantin Matchev$^{1}$
\\
$^{1}$Department of Physics and Astronomy, University of Alabama, Tuscaloosa, AL, 35487, USA\\
$^{2}$Department of Physics and Astronomy ‘Galileo Galilei’, University of Padova, Vicolo dell’Osservatorio 3, I-35122 Padova, Italy
}

\date{Accepted XXX. Received YYY; in original form ZZZ}

\begin{document}
\label{firstpage}
\pagerange{\pageref{firstpage}--\pageref{lastpage}}
\maketitle

\begin{abstract}
The rapid expansion of exoplanet atmospheric observations and the proliferation of a wide range of specialized modeling tools has created a need for flexible, accessible, and user-friendly workflows. Transmission spectroscopy, in particular, has become a key technique for probing atmospheric composition of transiting exoplanets. The analyses of these data require the combination of archival queries, literature search, the use of radiative transfer models, and Bayesian retrieval frameworks, each demanding specialized expertise.
Modern large language models (LLMs) enable the coordinated execution of complex, multi-step tasks by AI agents with tool integration, structured prompts, and iterative reasoning.
In this study we present ASTER, an Agentic Science Toolkit for Exoplanet Research. ASTER is an orchestration framework that brings LLM capability to the exoplanetary community by enabling LLM-driven interaction with integrated domain-specific tools, workflow planning and management, and support for common data analysis tasks.
Currently ASTER incorporates tools for downloading planetary parameters and observational datasets from the NASA Exoplanet Archive, as well as the generation of transit spectra from the TauREx radiative transfer model, and the completion of Bayesian retrieval of planetary parameters within the TauREx framework. 
Beyond tool integration, the agent assists users by proposing alternative modeling approaches, reporting potential issues and suggesting solutions, as well as higher-level interpretations. We demonstrate ASTER's workflow through a complete case study of WASP-39b, performing multiple retrievals using observational data from different sources available on the archive.
The agent efficiently transitions between datasets, generates appropriate forward model spectra and performs retrievals that recover atmospheric parameters reported in the literature.
ASTER provides a unified platform for the characterization of exoplanet atmospheres. 
Ongoing development and community contributions will continue expanding ASTER's capabilities toward broader applications in exoplanet research.
\end{abstract}

\section{Introduction}

The characterization of exoplanet atmospheres has rapidly evolved over the past decade, driven by increasingly precise observations and analysis frameworks. Transmission spectroscopy, in particular, has become a key technique for probing atmospheric composition of transiting exoplanets \citep{Charbonneau_2000,Seager_2000,Burrows_2014,Sing_2015,Tsiaras_2018,Edwards_2023}. Despite this progress, the analysis of transmission spectroscopy data remains a technically demanding and complex process \citep{Barstow_2020,Macdonald_2023}. A complete analysis typically requires the coordinated execution of multiple steps, including archival queries for stellar and planetary parameters, accessing observational datasets, forward modeling of synthetic spectra, and Bayesian retrieval analyses. These steps are managed manually, requiring substantial time and expertise to coordinate tools, interpret outputs, and diagnose failures at any stage of the workflow.

Recent advances in large language models (LLMs) offer new opportunities for managing such complex scientific pipelines \citep{NEURIPS2020_brown,yao2023react,schick2023toolformerlanguagemodelsteach,Zhang2025}. LLMs have progressed from passive text-generation systems into general-purpose computational interfaces capable of planning, reasoning, and interacting with external tools through structured prompts and function calls. When embedded within agentic frameworks, these models can coordinate multi-step workflows, and assist users in navigating large computational systems \citep{NEURIPS2023_yao,Plaat_2025}. Rather than replacing established scientific methods, such systems have the potential to orchestrate existing tools and increase efficiency, reproducibility, and accessibility.

Exoplanet atmospheric analysis provides a well-suited field for the development of agentic scientific workflows. Modern studies rely on complex computational pipelines that combine database queries, radiative-transfer modeling, and statistical inference, all of which must be coordinated and reproducible. In addition, machine learning is being increasingly used to help with the data simulation and parameter retrievals \citep{Zingales_2018,Marquez2018,Soboczenski2018,Nixon2020,Yip_2021,Ardevol2022,Yip2022,2023ApJ...958..106F,Eyup_2023,2023arXiv230909337A,Lueber_2025,2025arXiv250804982F,2026arXiv260102324R}. Yet existing pipelines often remain fragmented across notebooks and software packages, limiting reuse and increasing the barrier to entry for new researchers. At the same time, the rapid growth of exoplanet data from current and upcoming facilities accentuates the need for flexible and user-guided analysis frameworks (an up-to-date count of discovered exoplanets is available here\footnote{\hyperlink{https://exoplanetarchive.ipac.caltech.edu/}{https://exoplanetarchive.ipac.caltech.edu/}}).

In this work, we introduce ASTER (Agentic Science Toolkit for Exoplanet Research), an AI agent equipped with tools designed specifically for exoplanet atmospheric analysis. ASTER integrates LLM-based planning with exoplanet tools to support end-to-end transmission spectroscopy analyses. The current set of available tools enables: (i) automated download of planetary and stellar parameters from the NASA exoplanet archive \citep{Akeson_2013_exopl_archive,christiansen2025nasaexoplanetarchive}, (ii) generation of forward transmission spectra using the TauREx framework \citep{Waldmann_2015,Al_Refaie_2021}, (iii)  querying of publicly available transmission spectroscopy datasets from the NASA exoplanet archive, and (iv) Bayesian atmospheric retrieval using TauREx, for constraining atmospheric parameters from these observations. These tools are accessed through a structured interface that enables reliable coordination and reproducible execution \citep{roman2026orchestral}.

The agent functions as a workflow coordinator or assistant that executes tools, reads and interprets intermediate tool outputs, executes additional tools, takes steps to correct errors should they occur, summarizes or displays findings, and proposes next steps (such as alternative modeling assumptions, parameterizations, or data selections). The agent handles the technical implementation details while leaving scientific judgment and decision-making in the hands of the researcher. In this sense, ASTER operates not only as a tool executor, but as a scientific assistant that can identify potential issues, suggest improvements, and help contextualize results within the broader modeling landscape.

ASTER provides a foundation for adaptive exoplanet research workflows. While the present implementation focuses on transmission spectroscopy, the framework is designed to expand naturally to other areas of exoplanet science through community-driven tool development. More broadly, ASTER illustrates how agentic AI systems can enhance scientific reasoning and workflow management without displacing human expertise in the era of increasingly complex data and models.

The paper is organized as follows: 
In Section \ref{sec:methods}, we introduce the methodology of ASTER. We first describe the Orchestral AI framework \citep{roman2026orchestral}, including security hooks, cost tracking, and the set of built-in tools provided by Orchestral. 
In Section \ref{sec:tool_implementation}, we focus on the implementation of domain-specific tools within ASTER. We describe how tools are developed and integrated into the Orchestral framework. We then present the set of ASTER tools designed for exoplanet atmospheric analysis, which includes archive queries, forward modeling, data access, and retrieval workflows.
In Section \ref{sec:conv_examples}, we present example interactions with ASTER to illustrate how the agent responds to scientific questions. These examples are intended to present the agent’s reasoning process, tool usage, and interaction style, providing readers with a concrete understanding of how ASTER supports exploratory and guided analysis.
In Section \ref{sec:results}, we apply ASTER to the practical case of WASP-39b and discuss the results of the analysis. Using multiple publicly available transmission spectroscopy datasets, we demonstrate an end-to-end workflow.
Finally, Section \ref{sec:conclusion} summarizes our conclusions and discusses the limitations of the approach. We outline potential extensions of ASTER and future directions for agent-assisted exoplanet research.

\section{Methods}
\label{sec:methods}

\begin{figure}
    \centering
    \includegraphics[width=\linewidth]{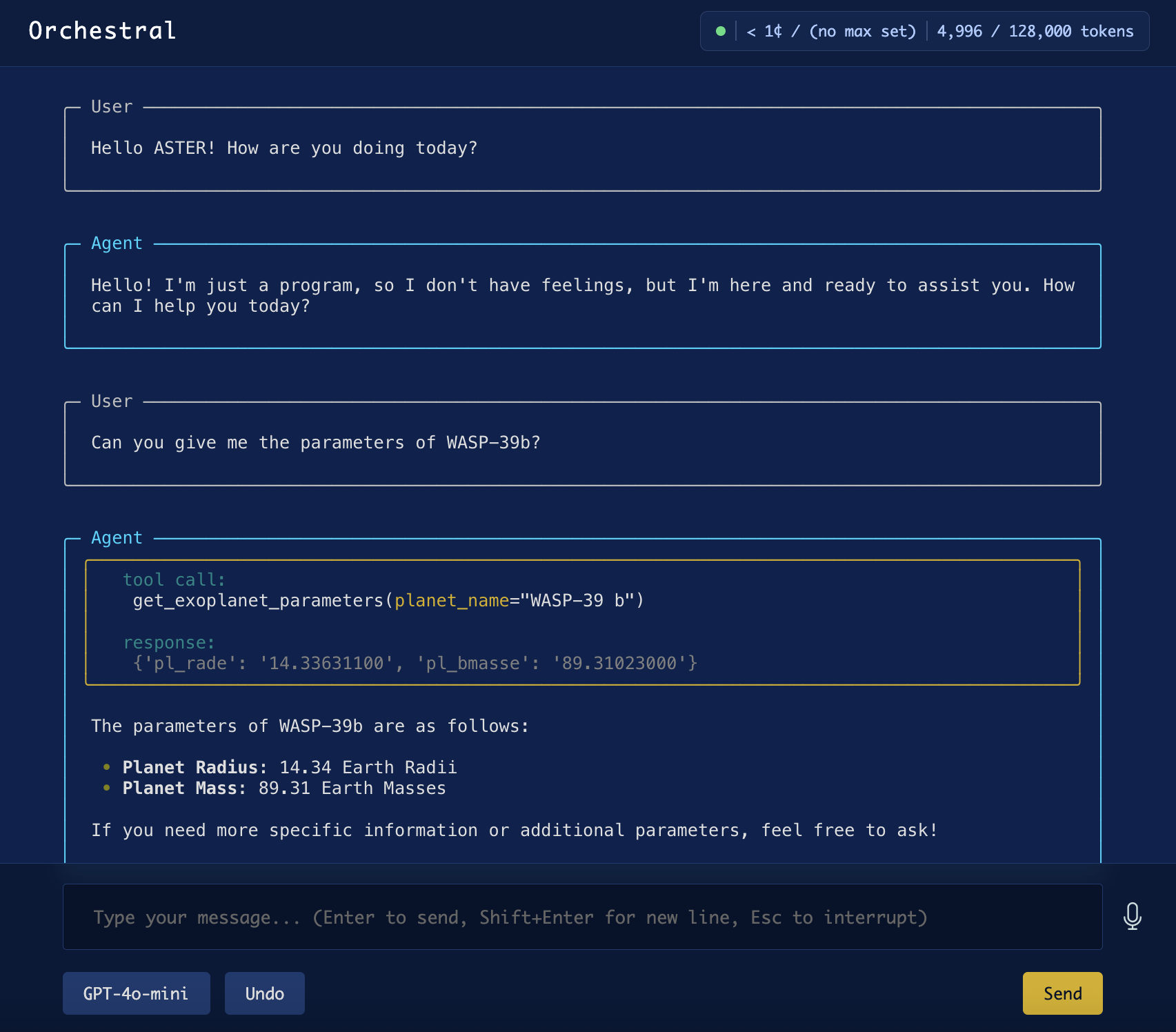}
    \caption{Example of the Orchestral interface. The LLM provider can be changed in the bottom left corner, below the space where users write their messages. The cost tracking can be seen in the top right corner.}
    \label{fig:interface}
\end{figure}

ASTER is implemented as an exoplanet-specific application of the Orchestral AI framework\footnote{\hyperlink{https://orchestral-ai.com/}{https://orchestral-ai.com/}} \citep{roman2026orchestral}. Orchestral provides the underlying agent infrastructure, handling the context management, tool schema construction and validation, tool execution, and controlled execution of multi-step workflows. The architecture of Orchestral is shown in Figure \ref{fig:architecture}, and an example of the user interface (UI) is shown in Figure \ref{fig:interface}. ASTER adopts this architecture and extends it with modular tools tailored to exoplanet atmospheric analysis. Other examples of AI agents built with the Orchestral AI framework applied to different domains in physics are presented in \cite{menzo_2025_heptapod}, \cite{knipfer2026}.

In this section and the following section, we describe the ASTER agent architecture, including an overview of the Orchestral AI framework, and ASTER's exoplanet tools.

\subsection{Orchestral overview}

Orchestral AI \citep{roman2026orchestral} is a Python framework designed for scientific applications of AI agents. It makes agent behavior transparent, debuggable, and portable across different LLM providers. Instead of hiding execution logic behind complex concepts, Orchestral keeps control flow explicit and easy to inspect.

One feature of Orchestral is provider agnosticism, meaning that the framework provides a unified interface to multiple LLM providers, such as OpenAI, Anthropic, or Google. This allows users to switch models without rewriting code, avoiding vendor lock-in. This is desirable in an institutional setting as universities and labs may require researchers to use a particular inference provider or local computing resources. Orchestral also uses a synchronous execution model, which makes agent behavior easier to debug, lowering the barrier to entry for domain experts in astrophysics unfamiliar with asynchronous programming.

\begin{figure}
    \centering
    \includegraphics[width=\linewidth]{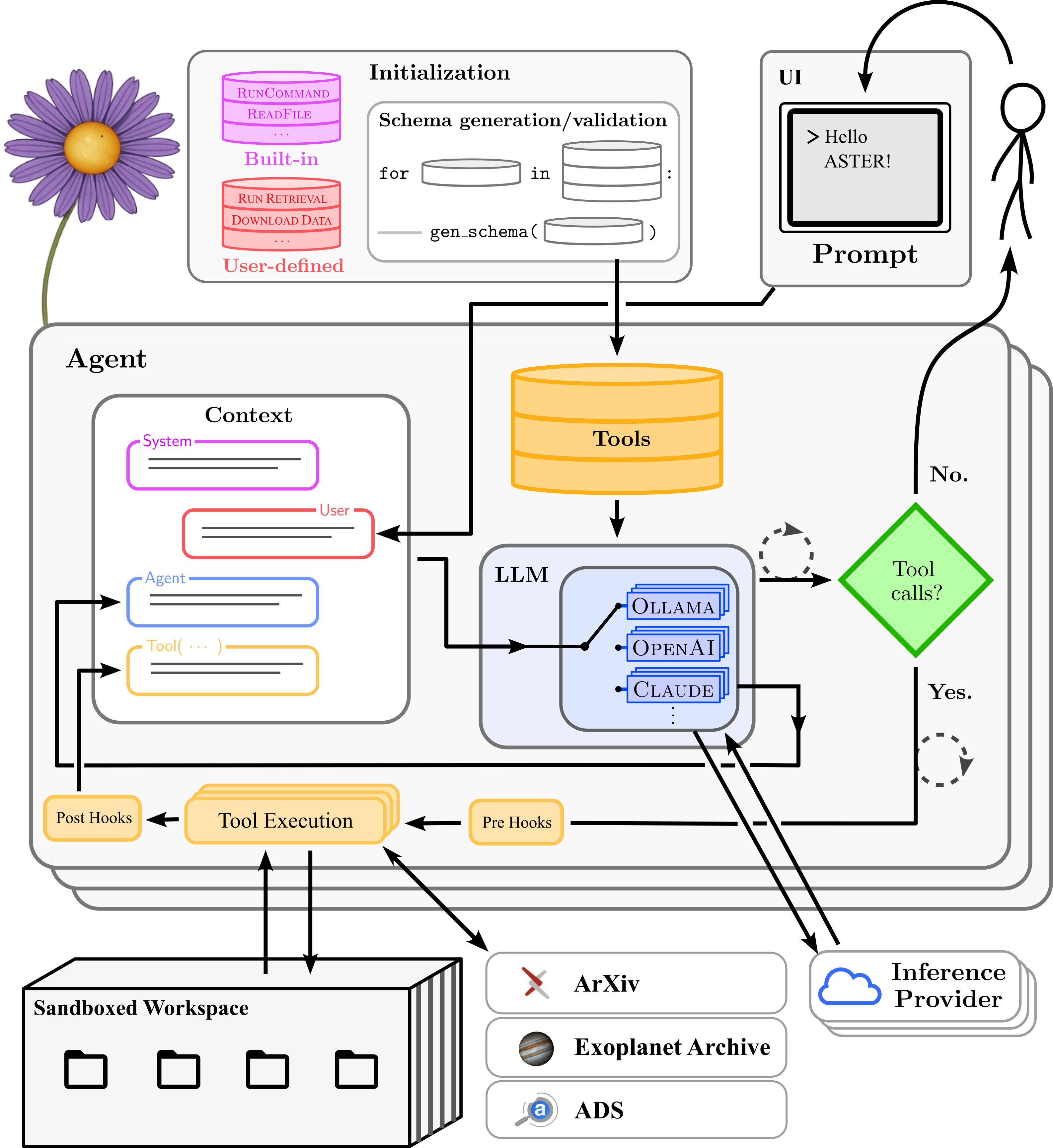}
    \caption{Schema of internal workings of the agent object. The agent comprises the LLM, the tools and the context. Figure adapted from \protect\cite{roman2026orchestral}.}
    \label{fig:architecture}
\end{figure}

In general, an AI agent is a system including the following core elements: An LLM, the ``context'', and a set of tools (see Figure \ref{fig:architecture}). AI agents may also include additional features such as longer-term memory, model selection mechanisms, security safeguards, and context management tools. 

The context is a record of all messages, tool calls, and tool outputs. In the Orchestral AI framework it also tracks associated metadata including token usage and cost. The context can be saved, loaded, and even continued with a different LLM provider. 
Orchestral automatically validates that tool calls and tool results match correctly. This prevents common errors that can occur when tool outputs are missing or misaligned with requests.

The LLM performs reasoning and selects actions based upon the current context. Typically it ``acts'' by generating tokens, which can be thought of as words (or parts of words). These may be in the form of natural language responses to the user (the typical behavior in chatbot tools such as ChatGPT) or in specially coded structures that indicate the use of tools or other special functions. If a user asks an agent equipped with a planetary data search tool the question ``what is the temperature of WASP 39'', the agent might immediately respond with ``\texttt{<tool> tool\_name: `get\_stellar\_info', parameters: \{star\_name: 'WASP-39'\}</tool>}'' and the system then detects and executes this pattern. Actual examples of ASTER tool calls are discussed in Section \ref{sec:conv_examples}. 
Defining tools in this way leverages ``in-context learning,'' a phenomenon in which LLMs can adapt their behavior based on information provided within the prompt, even when that information was not explicitly encountered during training \cite{brown2020languagemodelsfewshotlearners}.
We define a set of functions which carry out concrete actions, and present them to the agent along with descriptions of how and when to use them.

Tools form a bridge between the LLM and external software. In Orchestral, tools are easily defined as Python objects, the docstring and type hints are automatically processed into the ``tool schema'' language that the LLM expects. Orchestral also provides a Python decorator ``\texttt{@define\_tool}" which allows any Python function to be used as a tool.
One notable feature of tools is that they can return outputs in any format. Their outputs are inserted as strings into the conversation context alongside user messages. In the prior example, after querying the hypothetical \texttt{get\_stellar\_info} tool, the output might be ``\texttt{WASP-39: Temperature: $5400\pm150\, K$, Radius: $0.895±0.023\, R_\odot$, ...}''
In some cases the tool may return an error, for example, if the agent queried: \texttt{get\_stellar\_temp(`Death Star')} it might return: ``\texttt{Star not found, please check the stellarhosts database.}''


As illustrated in Figure \ref{fig:architecture}, the agent runs in an iterative loop. At each step, it reads the current context, asks the LLM to produce a response, checks whether tool calls are requested, executes those tools if present, and then updates the context with the results. This loop continues until no further tool calls are required or a stopping condition is reached. The LLM never executes code directly. All external actions happen through tools, which are executed outside the model in a controlled environment.\\

\textbf{Security hooks}

While modern commercial LLMs generally behave in predictable and desirable ways, they can sometimes take surprising and harmful actions. This is often the result of a ``misunderstanding'' of the user's intent on the part of the model, analogous to an ``honest mistake'' by human collaborators. However, a growing body of AI safety research has demonstrated that, under certain conditions, language models can also violate explicit instructions or engage in strategically harmful behavior, including deception or rule circumvention \cite{hubinger2024sleeperagentstrainingdeceptive, openai2024gpt4ocard}. Since these agents are equipped with tools granting them access to external systems, in some cases executing arbitrary terminal commands, it is imperative that guardrails be put in place to detect and block unwanted behavior.

Orchestral incorporates multiple layers of execution safety through a system of hooks that intercept tool calls before and after execution, enabling structured control over agent behavior and supporting safe, human-supervised workflows.

One key mechanism is user validation, implemented through a ``user-approval hook''. A simple approach is to require the user to approve all tool executions. While this guarantees safety, it is often impractical, as it interrupts the workflow and places a significant burden on the user. Heuristic rules provide a more efficient alternative by detecting and blocking clearly dangerous actions, such as destructive shell commands; however, they may miss more subtle risks and can be overly restrictive. 

A more flexible approach is to employ a separate LLM to evaluate proposed actions. In practice, specialized policy models can classify actions according to their safety, capturing more nuanced behavior than heuristics while remaining lightweight and efficient. However, such models are not perfectly reliable and must be used conservatively.

In Orchestral, these approaches are combined into a hybrid system. Heuristic checks first block clearly harmful actions, after which a safeguard LLM classifies remaining actions as safe, unsafe, or uncertain. Unsafe actions are blocked, while uncertain cases are escalated to the user for explicit approval. This layered approach balances safety, usability, and flexibility without unnecessarily interrupting routine workflows. \cite{roman2026orchestral}

These mechanisms ensure a strict separation between reasoning and execution: the language model proposes actions, while the orchestration layer enforces safety and validation.\\

\textbf{Cost tracking}

Orchestral includes built-in cost and token-usage tracking at the framework level. Every interaction with an LLM provider records token counts and associated costs, which are accumulated within the agent’s context object. This information persists across turns and tool calls, allowing users to monitor computational cost in real time. Because the context is provider-agnostic, cost tracking remains consistent even when switching between different models or providers within the same workflow. 
The Orchestral UI also supports setting a cost ceiling which will interrupt execution if reached.
This enables cost-aware experimentation, such as using lower-cost or local models for exploratory reasoning, followed by higher-capability models for final execution. These features are particularly relevant for long-running or iterative scientific analyses, where uncontrolled LLM usage could otherwise become quickly expensive.\\

\textbf{Conversations history}

Orchestral manages conversation history through the context object that encapsulates and validates the complete sequence of messages exchanged between the user, the LLM, and any external tools. This history includes prompts, model responses, tool calls, and tool outputs, ensuring that each step in the reasoning process is preserved and logically consistent. The system automatically verifies that tool calls and their results match correctly, and maintains provider-agnostic encoding so conversations can be saved and resumed across different LLM providers. 
This persistent conversational history is particularly valuable in science for bookkeeping and retracing a train of thoughts.\\

\textbf{Built-in tool set}\\

\textbf{Filesystem:}
\begin{itemize}
\addtolength{\itemindent}{1cm}
    \item \texttt{ReadFileTool}, reads files with character encoding detection, line ranges, and syntax highlighting for code and structured files. This tool is well-suited for inspecting code, configuration files, or large text documents efficiently.
    \item \texttt{WriteFileTool}, writes content to files with convenient features such as automatic backup and directory creation.
    \item \texttt{EditFileTool}, performs edits to existing files by applying only the necessary modifications. 
    \item \texttt{ListDirectoryTool}, recursively lists directory contents with optional filtering. This tool is useful for exploring project structures or preparing inputs for batch operations.
    \item \texttt{FileSearchTool}, executes regular-expression–based searches across files, returning matching lines along with surrounding context.
    \item \texttt{FindFilesTool}, locates files using glob-style patterns (e.g., **/*.py, *.md).\\
\end{itemize}

\textbf{Execution:}
\begin{itemize}
\addtolength{\itemindent}{1cm}
    \item \texttt{RunCommandTool}, provides access to a persistent shell session where the working directory, environment variables, and state are maintained across commands.
    \item \texttt{RunPythonTool}, executes Python code directly with a pre-loaded set of commonly used libraries. Suitable for data analysis, scripting, or validating logic without needing external setup.\\
\end{itemize}

\textbf{Web \& research:}
\begin{itemize}
\addtolength{\itemindent}{1cm}
    \item \texttt{WebSearchTool}, performs web searches and extracts structured results.\\
\end{itemize}

\textbf{Utilities:}
\begin{itemize}
\addtolength{\itemindent}{1cm}
    \item \texttt{TodoRead/TodoWrite}, enables task tracking directly within the conversation.
    \item \texttt{DisplayImageTool}, renders images in supported interfaces, allowing visual outputs to be displayed inline for clearer communication and analysis.
\end{itemize}

\section{Tool implementation}
\label{sec:tool_implementation}

We now turn to the practical question of how domain-specific functionality is implemented. As described earlier, all interactions between the agent and external software are accomplished through tools. Tools define the actions an agent is able to perform and form the building blocks from which the scientific workflow is constructed. This section describes how tools are defined and used within the Orchestral framework, with a focus on ASTER tools for exoplanet applications.

\subsection{Getting started}

ASTER builds on top of the Orchestral framework and requires additional dependencies specific to exoplanet atmospheric analysis. These requirements are maintained separately and are documented at \hyperlink{https://github.com/emipanek/aster}{https://github.com/emipanek/aster}.
Users should ensure that all ASTER-specific dependencies are installed before running or developing new tools.

\subsection{The role of tools}

The interaction between the agent and the tools follows a loop reasoning and execution pattern:
\begin{enumerate}
\addtolength{\itemindent}{6mm}
    \item The agent receives the current context.
    \item The language model responds, possibly invoking tools.
    \item If tool calls are not present, the response is passed to the user.
    \item If tool calls are present, the arguments are validated.
    \item The tool call passes though all pre-call hooks for approval.
    \item If approved, the tool is executed.
    \item The tool's output is passed though all post-call hooks.
    \item The processed output is added to the agent's context.
    \item Repeat.
\end{enumerate}

\subsection{How to write a tool}

Orchestral supports two primary ways of defining tools:
\begin{itemize}
\addtolength{\itemindent}{1cm}
    \item Function-based tools, using a decorator
    \item Class-based tools, for stateful or more complex behavior
\end{itemize}

In both cases, tools are written in standard Python syntax and rely on type hinting and docstrings to define their functioning. From these comments, Orchestral generates a universal tool object that can be used with any LLM. 

When using the function-based tool definition, the function signature represents the parameters the agent is allowed to control. The variable names will be visible to the agent, and it will decide based upon the context and the tool description (defined by the docstring) how to use these arguments.
Thus, clear and descriptive variable names and good documentation are strongly encouraged, as they inform the agent's decision-making process. 
When a tool is called, its code is executed once. The tool may call external libraries, run scientific models, or interact with data on disk. Tools should avoid side effects beyond what is necessary for their task and should always produce a well-defined output. 
Errors should be clearly passed to the agent along with relevant suggestions of how to remedy them, allowing the agent to detect failures and respond appropriately.

\subsection{Tool example}

Having described the general structure and behavior of tools, we now illustrate these concepts with a concrete example. The following example demonstrates how to define a simple tool, highlighting how inputs are specified, how execution is performed, and how results are returned to the agent.

\begin{strip}
\captionof{listing}{Example of the implementation of the ``GetExoplanetParameter'' tool. All of the documentation text in this example is captured and included as descriptions of the tool and arguments in the ``tool schema'' that is visible to the agent.}
\label{lst:python_function}
\begin{minted}[fontsize=\small,breaklines,breakanywhere,linenos]{python}
class GetExoplanetParameters(BaseTool):
    """
    Get exoplanet parameters from NASA Exoplanet Archive via TAP service.
    
    This tool allows programmatic access to exoplanet data from the NASA Exoplanet Archive.
    Uses the "pscomppars" table by default, which provides composite parameters for confirmed
    exoplanets combining data from multiple sources. Note that parameters may not be self-consistent
    as they come from different sources.
    
    IMPORTANT: Users should cite the archive's DOI if they use any datasets for research.
    """
    
    planet_name: str = RuntimeField(
        description="Name of the exoplanet (e.g., 'Kepler-10 b'), must match exactly with the archive."
    )
    columns: list = RuntimeField(
        default=["pl_radj", "pl_bmassj", "pl_eqt"],
        description="""List of parameter columns to retrieve. Common columns include:
        - pl_name: Planet name
        - hostname: Stellar name
        - pl_orbper: Orbital Period [days]
        - pl_orbsmax: Semi-major Axis [AU]
        - pl_rade: Planet Radius [Earth Radii]
        - pl_radj: Planet Radius [Jupiter Radii]
        - pl_bmasse: Planet Mass [Earth Masses]
        - pl_bmassj: Planet Mass [Jupiter Masses]
        - pl_eqt: Planet Equilibrium Temperature [K]
        - st_rad: Star Radius [Solar Radii]
        - st_teff: Star Temperature [K]
        - st_mass: Star Mass [Solar Masses]
        - st_met: Stellar Metallicity [dex]
        - st_logg: Stellar surface gravity [log10(cm/s^2)]
    
        Full list at: https://exoplanetarchive.ipac.caltech.edu/docs/API_PS_columns.html"""
    )
    ...
\end{minted}

\captionof{listing}{Abbreviated example code to define and use the ASTER agent.}
\label{lst:tool_implementation}
\begin{minted}[fontsize=\small,breaklines,breakanywhere,linenos]{python}
from orchestral import Agent
# Other Orchestral imports ...

from aster_toolkit import (
    RunTaurexModelTool,
    SetTaurexPaths,
    SimulateTaurexRetrieval,
    GetExoplanetParameters,
    DownloadDataset
)

# Set workspace directory
base_directory = 'workspace'

# Construct list of tools
tools = [
    # File and command tools
    RunCommandTool(base_directory=base_directory, persistent=True),
    WriteFileTool(base_directory=base_directory),
    ReadFileTool(base_directory=base_directory, show_line_numbers=True),
    EditFileTool(base_directory=base_directory),
    FileSearchTool(base_directory=base_directory),
    WebSearchTool(),
    TodoRead(),
    TodoWrite(),
    DisplayImageTool,

    # TauREx modeling tools
    SetTaurexPaths,
    RunTaurexModelTool(base_directory=base_directory),
    SimulateTaurexRetrieval(base_directory=base_directory),

    # Data acquisition tools
    GetExoplanetParameters(),
    DownloadDataset(base_directory=base_directory)
]

agent = Agent(
    # llm=Claude(),
    llm=GPT(model="gpt-4.1-mini"),
    tools=tools,
    tool_hooks=[UserApprovalHook()],
    system_prompt=system_prompt
)

# Launch the UI
from orchestral.ui.app import server as app_server
app_server.run_server(agent, host='localhost', port=8000)

\end{minted}
\end{strip}

Listing \ref{lst:tool_implementation} illustrates the general structure and behavior of an Orchestral tool within ASTER. While individual tools differ in their scientific purpose, they all follow the same design principles: explicit instructions explaining how and when to use the tool, clearly defined inputs, and structured outputs that can be composed into larger workflows.

\subsection{Overview of available ASTER tools}

This subsection summarizes the exoplanet-specific tools implemented within ASTER at the time of writing, which extend the built-in Orchestral tool set described earlier. These exoplanet tools provide the functionality needed for end-to-end transmission spectroscopy analyses, including database queries, forward-model generation, data access, and retrieval execution.

\subsubsection{Downloading planetary systems parameters}

This tool uses the NASA exoplanet archive's\footnote{\href{https://exoplanetarchive.ipac.caltech.edu/}{https://exoplanetarchive.ipac.caltech.edu/}} TAP (Table Access Protocol) service to directly download planetary parameters (see \cite{Akeson_2013_exopl_archive} and \cite{christiansen2025nasaexoplanetarchive} for further details on the functionality of NASA Exoplanet archive). This tool is the example presented in the tool example of Listing \ref{lst:python_function}. The most important parameters for our transmission spectroscopy example are the planet parameters, such as planet radius, planet mass, or equilibrium temperature, as well as the star parameters: stellar radius, stellar mass and effective stellar temperature. Users can request any available parameter set from the Planetary Systems (ps) or Planetary Systems Composite Parameters (pscomppars) tables. The default is to use pscompars table. In the Planetary Systems table, one exoplanet can have multiple sets of parameters from various papers, each table row contains one parameter set from one paper, leaving empty cells when values are not provided. This is the reason why we are using the composite table, which provides a single set of composite parameters for each confirmed exoplanet, combining data from multiple sources. One downside of using this table is that the parameters are not necessarily self-consistent, as they are taken from different sources. This tool allows us to access up-to-date values in CSV format. 
\footnote{An interesting note is that the agent is really flexible regarding the planet name, it will still print the right parameters if it is being asked for wasp 39b, WASP-39b or even waps-39B.}

\subsubsection{Running a forward model}

This tool uses TauREx \citep{Waldmann_2015,Al_Refaie_2021} to simulate theoretical transmission spectra. TauREx \citep{Waldmann_2015,Al_Refaie_2021} is a radiative transfer code that calculates spectra of exoplanet atmospheres, using a plane-parallel approximation. In order to compute how an atmosphere absorbs and transmits light, TauREx needs opacity and Collision Induced Absorption (CIA) files. Opacity files store the wavelength dependent absorption properties of molecules and CIA files describe absorption that occurs not because of individual molecules, but because pairs of molecules collide. For this example we downloaded our opacity files from the Exomol\footnote{\hyperlink{https://www.exomol.com/data/molecules/}{https://www.exomol.com/data/molecules/}} website and our CIA files from HITRAN\footnote{\hyperlink{https://hitran.org/cia/}{https://hitran.org/cia/}}. The references for the molecular cross-section data used in this study are listed in Table \ref{tab:cross_sections}. The tool starts by checking if the opacity and CIA file paths required by TauREx are set, and asks the user to provide them if not. Once the paths are confirmed, the tool is generating a transmission model using the parameters previously downloaded. The agent can also call the two tools at once, if the user directly asks for a model for a specific planet in a single message. For this study, the agent builds an isothermal temperature profile. 
This tool overall ensures that the TauREx environment is set properly and facilitates easy modeling of spectra for different exoplanets.

\begin{table}
\centering
\begin{tabular}{ll}
\hline
\textbf{Molecule} & \textbf{Reference} \\
\hline
H$_2$O & \cite{Polyansky_2018_H2O,Chubb_2020_exomol} \\
CO & \cite{Li_2015_CO,Chubb_2020_exomol} \\
CO$_2$ & \cite{Yurchenko_2025_CO2} \\
CH$_4$ & \cite{Yurchenko_2024_CH4} \\
NH$_3$ & \cite{Coles_2019_NH3,Chubb_2020_exomol} \\
HCN & \cite{Barber_2013_HCN,Chubb_2020_exomol} \\
H$_2$S & \cite{Azzam_2016_H2S,Chubb_2020_exomol} \\
C$_2$H$_2$ & \cite{Chubb_2020_C2H2,Chubb_2020_exomol} \\
Na & \cite{Allard_2019_na,Chubb_2020_exomol}  \\
K & \cite{Allard_2016_K,Chubb_2020_exomol} \\
TiO & \cite{McKemmish_2019_Tio,McKemmish_2024_Tio} \\
VO & \cite{Bowesman_2024_VO} \\
\hline
\end{tabular}
\caption{Sources of cross-section data used in this work, downloaded from the Exomol website.}
\label{tab:cross_sections}
\end{table}

\subsubsection{Downloading available transmission spectroscopy data}

This tool downloads transmission spectroscopy datasets from the NASA Exoplanet Archive (see \cite{Akeson_2013_exopl_archive} and \cite{christiansen2025nasaexoplanetarchive} for further details on the functionality of NASA Exoplanet archive), using the generation of the necessary \texttt{wget} commands. The user can ask to download a set of parameters and the agent will walk the user through the following process: 

\begin{enumerate}
\addtolength{\itemindent}{1cm}
    \item Please go to \hyperlink{https://exoplanetarchive.ipac.caltech.edu/cgi-bin/atmospheres/nph-firefly?atmospheres}{https://exoplanetarchive.ipac.caltech.edu/cgi-bin/atmospheres/nph-firefly?atmospheres} 
    \item Use the filters on the website to select the desired planet and choose the instruments or specific observations you want.
    \item Select the spectra by checking the boxes.
    \item Click on ``Download all checked spectra''.
    \item The website will generate \texttt{wget} commands or provide a URL to a page with \texttt{wget} commands.
    \item Please provide me either the \texttt{wget} commands as text, the URL to the \texttt{wget} page, or if you saved the \texttt{wget} commands to a file, upload the file or give me the file path.\\
\end{enumerate}

The agent can then download and process the selected datasets. This process allows users to access any public dataset available on the NASA Exoplanet Archive without needing to manually reformat the data for retrieval analysis. The tool cannot be fully automated because the generated \texttt{wget} commands are not persistent over time.
Once the path to the \texttt{wget} commands is provided, the agent executes those commands to download IPAC tables and saves them as per-planet CSV files. The agent then reads the CSV files, extracts the relevant spectral information, and produces two outputs: a data file (spectrum.dat) and a metadata file (metadata.json).
The data file contains three columns: wavelength (in microns), transit depth, and the error on the transit depth. The metadata file includes the authors of the dataset’s first publication, a URL to the associated paper, the observing facility and instruments used, and the units corresponding to each column in the data file.
Direct use of the NASA Exoplanet Archive TAP system is not viable for this task because the legacy 'transitspec' table has been retired and is no longer being updated after the summer of 2023. The archive explicitly directs users to the newer Atmospheric Spectroscopy Table instead, which is not retrievable through TAP in the same way.

\subsubsection{Running a retrieval}

This next tool runs a TauREx \citep{Waldmann_2015,Al_Refaie_2021} retrieval starting from an observed transmission spectrum and a simple forward model. The user can either provide the path to an observation file or let the agent download a dataset using the tool described previously. TauREx supports multiple retrieval modes: reduced, full, and equilibrium. The reduced mode uses a predefined set of common molecules: H$_2$O, CH$_4$, CO$_2$, CO and NH$_3$. This is the simplest configuration and is therefore used as the default when no mode is explicitly specified by the user. The full mode allows the user to define a custom set of molecules. The user can also ask the agent to use every molecule available, in which case the agent inspects the available cross-section files in the path specified earlier and uses the complete set of molecules. The user also defines an initial mixing ratio for each molecule, as the starting point before the retrieval begins. In both the reduced and full modes the fitted parameters are the mixing ratios of each molecules. The equilibrium mode relies on ACE \citep{Agundez_2012,Agundez_2020} to compute the thermochemical equilibrium abundances by minimizing the Gibbs free energy. In this case the fitted parameters are the metallicity and C/O ratio.
TauREx currently supports the following optimizers: Nestle \citep{Skilling_2004_nestle,Mukherjee_2006_nestle, Shaw_2007_nestle,Feroz_2009}, MultiNest \citep{Feroz_2009,Buchner_2014_multinest_citation}, and UltraNest \citep{buchner_2021_ultranest}. All three rely on nested sampling, a Bayesian inference method that explores the parameter space and simultaneously estimates the Bayesian evidence \citep{Buchner_2014_NS_test,Buchner_2019}. Nested sampling is well-suited for multi-modal posterior distributions and problems with parameter degeneracies because it explores parameter space globally rather than locally. While all three are based on nested sampling, they differ in their practical use cases. Nestle is lightweight and faster, making it more appropriate for simple, low-dimensional problems. MultiNest is well-established and commonly applied in exoplanets analysis. UltraNest is a more recent implementation that provides improved numerical stability and better performance for high-dimensional parameter spaces.

\subsubsection{Plotting the results}

This tool generates corner plots from the posterior of a retrieval run, using the corner library \citep{Foreman-Mackey2016}. It offers an easy way to see the parameter distributions and their correlations. Corner plots are common in retrieval studies because they reveal the overall structure of the solution space. 

The agent can automatically generate simple corner plots without user intervention, returning them alongside the retrieval results. A basic corner plot is also produced and stored when the retrieval tool is executed. This dedicated plotting tool extends that functionality by providing sensible defaults and configurable arguments for commonly adjusted formatting options. It is particularly useful for comparing multiple retrieval runs, as it can overlay posteriors from different results within a single corner plot, facilitating direct comparison. For more specialized visualizations, the code-based approach discussed in Section \ref{sec:skills} can be used.


\subsection{Sequential tool execution}

These tools get used in combination to accomplish complicated tasks, sometimes without the user explicitly asking. For example, if you ask for a transmission spectrum of WASP-39b, the agent knows that entails first searching for the parameters and then running the forward model tool. 
The agent decomposes the request into a sequence of subtasks, selects the appropriate tools, and executes them in the correct order. This behavior highlights the agent's ability to perform implicit planning, allowing it to autonomously execute multi-step scientific tasks. 

\subsection{Tools vs. Skills}
\label{sec:skills}
A recurring idea in recent work on coding agents is that code provides a near-universal interface to computation: given sufficient capability and access to a terminal environment, an agent can in principle implement arbitrary functionality through code generation and execution. This perspective, which became increasingly prominent in 2024–2025, has led to the view that instead of building specialized tools, one can equip agents with ``skills'', typically expressed as natural language instructions or reusable code snippets provided in context.
In this paradigm, a skill may take the form of a structured prompt, markdown file, or retrieval mechanism that injects task-specific guidance into the agent’s context. In contrast, a tool is a predefined function with a fixed interface that executes a specific operation.

We argue that both approaches have complementary strengths. On one hand, if an agent has access to a sufficiently expressive environment such as a terminal, many tasks can in principle be accomplished through direct code generation. This suggests that not all functionality must be exposed as explicit tools.
On the other hand, tools are valuable when reliability, safety, or structure is required. For example, if agents consistently struggle with a recurring task, encapsulating that functionality in a tool can improve robustness.
Tools can also enforce constraints or maintain state in ways that are difficult to achieve through prompting alone. For example, in Orchestral the file read and edit system tracks which files an agent has read and blocks blind modifications to unseen files, thereby reducing the risk of unintended changes.
Other tasks exhibit a high degree of variability and open-ended structure that is difficult to capture with a fixed function interface. For instance, generating a scientific plot involves a large design space of stylistic and semantic choices, making it cumbersome to parameterize as a tool. In such cases, requiring the agent to directly write code is often more natural and efficient.
For such open-ended tasks, skills provide a flexible, user-friendly mechanism for guiding agent behavior. They also allow non-technical users to encode preferences, domain knowledge, and heuristics in natural language without needing to define rigid interfaces.

In our own experiments, we initially implemented a dedicated tool for generating corner plots of posterior distributions, but later found that we preferred the agent to use its built-in coding tools to write the code directly, guided by a skill describing best practices and stylistic preferences.
Conversely, in our testing, we observed that the agent required trial and error to correctly configure TauREx and, in some cases, hallucinated a nonexistent TauREx \texttt{run\_transit()} function. We contend that this is a case where a dedicated tool such as \texttt{RunTaurexModel} greatly improves reliability and efficiency.

\section{Conversation example}
\label{sec:conv_examples}

To illustrate the range of tasks supported by the agent, we present here an example conversation between a user and the agent. The user input is shown in the ``User'' black boxes, the agent output is shown in the ``Agent'' blue boxes, and the orange boxes indicate a tool call, where one can explicitly see the parameters chosen by the agent. This exchange demonstrates how different requests are formulated in natural language and how the agent interprets and executes them, from data acquisition to analysis and visualization. 

\begin{orchestralusermessage}
    Hey ASTER! How are you doing? Describe the available tools
\end{orchestralusermessage}

\begin{orchestralagentmessage}
    Hello! I'm here and ready to assist you with exoplanet atmospheric research using TauREx spectral modeling. \\
    Here's a description of the tools available to me:\\

    \begin{enumerate}
    \addtolength{\itemindent}{5mm}
      \item \textbf{Get Exoplanet Parameters}: Retrieve planet and host star parameters from the NASA Exoplanet Archive for specific exoplanets.
      \item \textbf{Download Dataset}: Download transit spectra of exoplanets from the NASA Exoplanet Archive using \texttt{wget} commands or URLs provided by the user.
      \item \textbf{Run TauREx Forward Model}: Generate synthetic forward model transmission spectra for exoplanets based on given parameters.
      \item \textbf{Run TauREx Atmospheric Retrieval}: Perform atmospheric retrievals by fitting model parameters to observed transmission spectra using nested sampling algorithms.
      \item \textbf{Set TauREx Paths}: Configure the paths to linelists opacity (xsec) and collision-induced absorption (CIA) files, which are required inputs for TauREx.
      \item \textbf{File Operations}: Read, write, and edit files within the workspace to manage configuration and data files.
      \item \textbf{Run Command}: Execute shell commands in the workspace environment.
      \item \textbf{Search Tools}: Search file contents using regex or perform web searches for additional information.
      \item \textbf{Display Images}: Show plots and figures generated during the analysis.
      \item \textbf{Todo List Management}: Keep track of tasks and progress in the current conversation.\\
    \end{enumerate}

    These tools allow me to assist you end-to-end in exoplanet atmospheric modeling, from data acquisition and forward modeling to atmospheric retrieval and analysis. How can I assist you today?

\end{orchestralagentmessage}

\begin{orchestralusermessage}
    Can you give me parameters of wasp39 b ?
\end{orchestralusermessage}

\begin{orchestralagentmessage}

    \begin{orchestraltoolmessage}{GetExoplanetParameters}
        \textcolor{teal}{tool call:} \\ 
        GetExoplanetParameters(\\
        \hspace*{1em}\textcolor{brown}{planet\_name}="WASP-39b", \\
        \hspace*{1em}\textcolor{brown}{columns}="['pl\_name', 'hostname', 'pl\_orbper', 'pl\_radj', 'pl\_orbsmax', 'pl\_bmassj', 'pl\_eqt', 'st\_rad', 'st\_teff']"\\
        )\\
        \textcolor{teal}{response:}
        
        \textcolor{gray}{
        Parameters for WASP-39b:\\}
        
          \textcolor{gray}{Planet Name: WASP-39b\\
          Star Name: WASP-39\\
          Orbital Period: 4.0553 days\\
          Semi-major Axis: 0.04828 AU\\
          Planet Radius: 1.279 RJup\\
          Planet Mass: 0.281 MJup\\
          Equilibrium Temp: 1166 K\\
          Star Radius: 0.939 Rsun\\
          Star Temp: 5485 K\\
          Star Mass: 0.913 Msun\\
          Stellar Metallicity: 0.01 dex\\
          Stellar log(g): 4.453 log10(cm/s$^2$)}
    \end{orchestraltoolmessage}

    Here are the parameters for WASP-39b:

    \begin{itemize}
    \addtolength{\itemindent}{3mm}
      \item Planet Name: WASP-39b
      \item Star Name: WASP-39
      \item Orbital Period: 4.0553 days
      \item Semi-major Axis: 0.04828 AU
      \item Planet Radius: 1.279 Jupiter radii
      \item Planet Mass: 0.281 Jupiter masses
      \item Equilibrium Temperature: 1166 K
      \item Star Radius: 0.939 Solar radii
      \item Star Temperature: 5485 K
      \item Star Mass: 0.913 Solar masses
      \item Stellar Metallicity: 0.01 dex
      \item Stellar surface gravity (log g): 4.453 log10(cm/s$^2$)\\
    \end{itemize}

    If you need any further information or want to proceed with modeling or retrieval using these parameters, just let me know!

\end{orchestralagentmessage}

\begin{orchestralusermessage}
    Can you make a forward model?
\end{orchestralusermessage}

\begin{orchestralagentmessage}

    \begin{orchestraltoolmessage}{RunCommand: \texttt{'ls linelists/'}}
        \textcolor{teal}{output:}\\
        Command: ls linelists/  \\
        Return Code: 1 \\
        Output: \\
        ls: linelists/: No such file or directory \\
    \end{orchestraltoolmessage}

    It seems the linelists directory needed for TauREx forward modeling is not present in the workspace. The linelists directory should contain the subdirectories 'xsec/' for opacity files and 'cia/' for collision-induced absorption files.\\

    Do you want me to guide you on how to download or set up the required linelists for TauREx?

\end{orchestralagentmessage}

\begin{orchestralusermessage}
    Here they are: "/Path/to/xsec" "/Path/to/cia\_hitran"
\end{orchestralusermessage}

\begin{figure}[h]
    \centering
    \includegraphics[width=\linewidth]{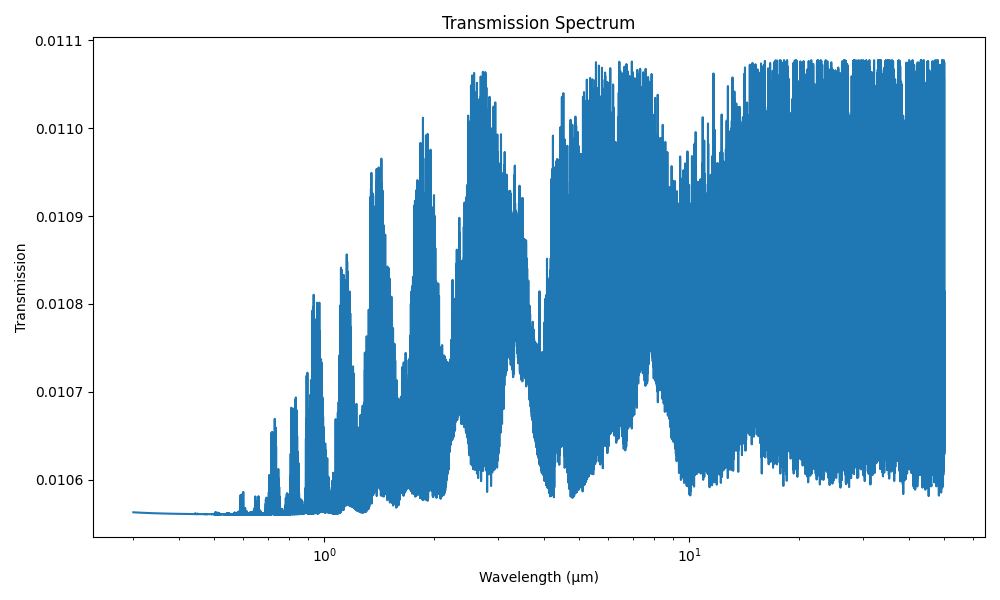}
    \caption{A plot generated, formatted, and displayed by the agent during the example conversation. It represents the transmission spectrum for WASP-39b, calculated by the agent and shown using the \texttt{DisplayImage} tool.}
    \label{fig:display_example}
\end{figure}

\begin{orchestralagentmessage}

    \begin{orchestraltoolmessage}{SetTaurexPaths}
        \textcolor{teal}{tool call:} \\ 
        SetTaurexPaths(\\
        \hspace*{1em}\textcolor{brown}{opacity\_path}="/path/to/xsec", \\
        \hspace*{1em}\textcolor{brown}{cia\_path}="/path/to/cia"\\
        )\\
        \textcolor{teal}{response:}
        
        \textcolor{gray}{Paths set successfully!}
        
    \end{orchestraltoolmessage}

    \begin{orchestraltoolmessage}{RunTaurexModel}
        \textcolor{teal}{tool call:} \\ 
        RunTaurexModel(\\
        \hspace*{1em}\textcolor{brown}{star\_radius}="0.939", \\
        \hspace*{1em}\textcolor{brown}{planet\_radius}="1.279", \\
        \hspace*{1em}\textcolor{brown}{planet\_mass}="0.281", \\
        \hspace*{1em}\textcolor{brown}{orbital\_period}="4.0553", \\
        \hspace*{1em}\textcolor{brown}{semi\_major\_axis}="0.04828", \\
        \hspace*{1em}\textcolor{brown}{planet\_temp}="1166", \\
        \hspace*{1em}\textcolor{brown}{atm\_min\_pressure}="0.001", \\
        \hspace*{1em}\textcolor{brown}{atm\_max\_pressure}="100.0", \\
        \hspace*{1em}\textcolor{brown}{filename}="wasp39b\_forward\_model"\\
        )\\
        \textcolor{teal}{response:}
        
        \textcolor{gray}{Taurex model run successfully with output file: wasp39b\_forward\_model\_spectrum.png}
    \end{orchestraltoolmessage}

    \begin{orchestraltoolmessage}{DisplayImage}

        \textcolor{teal}{image:}

        wasp39b\_forward\_model\_spectrum.png (Figure~\ref{fig:display_example})\\
    \end{orchestraltoolmessage}

    I have generated a forward model transmission spectrum for WASP-39b using the available parameters. Here is the plot of the synthetic spectrum. If you want, I can help with atmospheric retrievals, further analysis, or other tasks. Just let me know!

\end{orchestralagentmessage}

\begin{orchestralusermessage}
    Please download spectra for wasp39b with nirspec
\end{orchestralusermessage}

\begin{orchestralagentmessage}
    To download spectra for WASP-39b observed with NIRSpec, please follow these steps:\\

    \begin{enumerate}
    \addtolength{\itemindent}{1cm}
      \item Go to the NASA Exoplanet Archive atmospheres Firefly interface:  \hyperlink{https://exoplanetarchive.ipac.caltech.edu/cgi-bin/atmospheres/nph-firefly}{https://exoplanetarchive.ipac.caltech.edu/cgi-bin/atmospheres/nph-firefly}
      \item Use the filters to select planet: ``WASP-39b'' and instrument: ``NIRSpec''.
      \item Check the boxes next to the spectra you want to download.
      \item Click on ``Download all checked spectra''.
      \item The website will provide you with \texttt{wget} commands or a URL to a \texttt{wget} command page.\\
    \end{enumerate}

    Please upload or paste ONE of the following for me:

    \begin{itemize}
    \addtolength{\itemindent}{1cm}
      \item The \texttt{wget} commands text directly
      \item The URL to the \texttt{wget} page (starting with https://exoplanetarchive.ipac.caltech.edu/staging/)
      \item Or if you saved the \texttt{wget} commands to a file, upload that file\\
    \end{itemize}

    Then I can download and organize the spectra for you.

\end{orchestralagentmessage}

\begin{orchestralusermessage}
    https://exoplanetarchive.ipac.caltech.edu/work/TMP\_rnIuBD\_33318/ atmospheres/tab1/wget\_atmospheres.bat
\end{orchestralusermessage}

\begin{orchestralagentmessage}

    \begin{orchestraltoolmessage}{DownloadDataset}
        \textcolor{teal}{tool call:} \\ 
        DownloadDataset(\\
        \hspace*{1em}\textcolor{brown}{wget\_url}="https://exoplanetarchive.ipac.caltech.edu/work... \\
        \hspace*{1em}\textcolor{brown}{output\_dir}="spectra/WASP\_39\_b\_nirspec"\\
        )\\
    
        \textcolor{teal}{response:}

        \textcolor{gray}{Download complete! Query ID: query002}\\

        \textcolor{gray}{Planets processed: WASP\_39\_b\_3}\\

        \textcolor{gray}{Working files: download\_dataset\_tool/query002/
        Final spectra saved to: spectra/WASP\_39\_b\_nirspec/}\\

        \textcolor{gray}{Spectrum file paths:\\
          - spectra/WASP\_39\_b\_nirspec/WASP\_39\_b\_3/WASP\_39\_b \_3.11466\_4988\_1/spectrum.dat\\
          - spectra/WASP\_39\_b\_nirspec/WASP\_39\_b\_3/WASP\_39\_b \_3.11466\_5087\_1/spectrum.dat\\
          - spectra/WASP\_39\_b\_nirspec/WASP\_39\_b\_3/WASP\_39\_b \_3.11466\_5077\_1/spectrum.dat\\
          - spectra/WASP\_39\_b\_nirspec/WASP\_39\_b\_3/WASP\_39\_b \_3.11466\_5502\_6/spectrum.dat\\
          - spectra/WASP\_39\_b\_nirspec/WASP\_39\_b\_3/WASP\_39\_b \_3.11466\_4988\_2/spectrum.dat\\
          - spectra/WASP\_39\_b\_nirspec/WASP\_39\_b\_3/WASP\_39\_b \_3.11466\_4988\_4/spectrum.dat\\
          - spectra/WASP\_39\_b\_nirspec/WASP\_39\_b\_3/WASP\_39\_b \_3.11466\_4988\_3/spectrum.dat\\
          - spectra/WASP\_39\_b\_nirspec/WASP\_39\_b\_3/WASP\_39\_b \_3.11466\_5502\_4/spectrum.dat\\
          - spectra/WASP\_39\_b\_nirspec/WASP\_39\_b\_3/WASP\_39\_b \_3.11466\_5502\_5/spectrum.dat}\\
    \end{orchestraltoolmessage}

    The NIRSpec spectra for WASP-39b have been successfully downloaded and organized in the directory spectra/WASP\_39\_b\_nirspec/.

    There are multiple spectrum files available. If you want, I can help you analyze, visualize, or perform atmospheric retrievals on any of these datasets. Just let me know which specific spectrum file you are interested in or what you would like to do next.

\end{orchestralagentmessage}

\begin{orchestralusermessage}
    Can you now do a retrieval using the multinest optimizer and fitting temperature, planet radius and the molecular abundances? 
\end{orchestralusermessage}

\begin{figure}
    \centering
    \includegraphics[width=\linewidth]{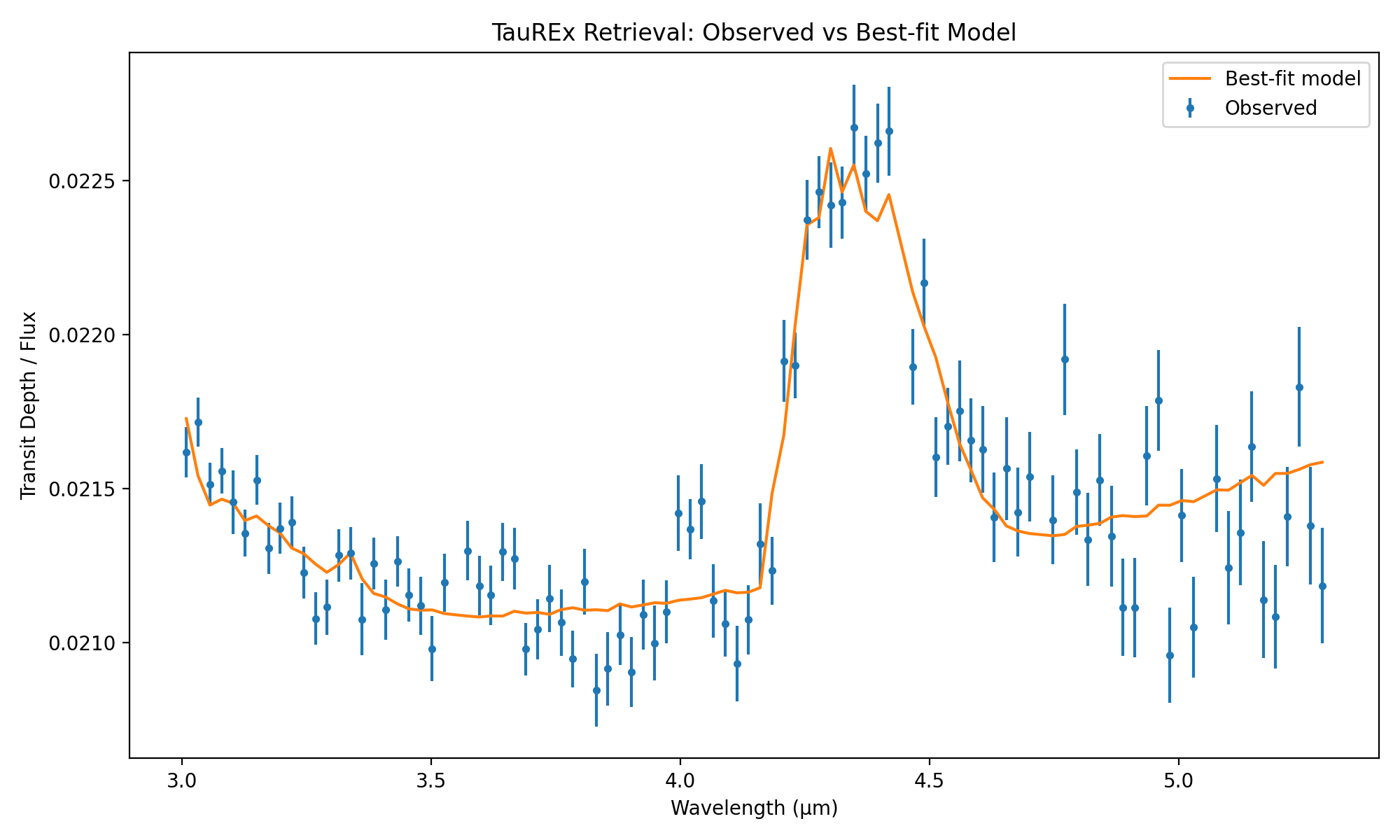}
    \caption{A plot generated, formatted, and displayed by the agent during the example conversation. It represents the best-fit model (orange curve) retrieved by the agent on WASP-39b NIRSpec observation (blue points).}
    \label{fig:fit_example}
\end{figure}

\begin{orchestralagentmessage}

    \begin{orchestraltoolmessage}{ReadFile: skills/retrieval\_best\_practices.md}
        \textcolor{teal}{output:}\\
        File: skills\_retrieval\_best\_practices.md (lines 1-20 of 180)\\
        
\begin{minted}{text}
File: skills/retrieval_best_practices.md 
(lines 1-50 of 180)
```markdown
1→# Retrieval Best Practices
2→
3→This skill provides guidance on setting up 
atmospheric retrievals for optimal performance 
and physically meaningful results.
4→
5→## Parameter Bounds
6→
7→### General Principles
8→
9→The bounds you set directly control:
10→1. **Runtime**: Wider bounds = longer 
retrieval time
11→2. **Physical plausibility**: Bounds should 
reflect realistic atmospheric conditions
12→3. **Convergence**: Overly wide bounds can 
make it harder to find solutions
13→
14→### Molecular Abundance Bounds
15→
16→**Chemical abundance limits (log10 mixing 
ratio)**:
17→- **Lower limit**: `-9` 
18→  - Abundances below this limit have 
essentially no measurable spectroscopic effect
19→  - Values below this are computationally 
wasteful and practically indistinguishable
20→  - **Never use bounds lower than -9**
...
```
\end{minted}
    \end{orchestraltoolmessage}

\begin{orchestraltoolmessage}{SimulateTaurexRetrieval}
        \textcolor{teal}{tool call:} \\ 
        SimulateTaurexRetrieval(\\
        \hspace*{1em}\textcolor{brown}{observation\_path}="download\_dataset\_tool/query001/...\\ 
        \hspace*{1em}\textcolor{brown}{fit\_params}="['planet\_radius', 'T', 'H2O', 'CH4', 'CO', 'CO2', 'NH3']", \\
        \hspace*{1em}\textcolor{brown}{optimizer}="multinest", \\
        \hspace*{1em}\textcolor{brown}{num\_live\_points}="100", \\
        \hspace*{1em}\textcolor{brown}{star\_radius}="0.939", \\
        \hspace*{1em}\textcolor{brown}{planet\_radius}="1.279", \\
        \hspace*{1em}\textcolor{brown}{planet\_mass}="0.281", \\
        \hspace*{1em}\textcolor{brown}{planet\_temp}="1166", \\
        \hspace*{1em}\textcolor{brown}{atm\_min\_pressure}="0.001", \\
        \hspace*{1em}\textcolor{brown}{atm\_max\_pressure}="100000", \\
        \hspace*{1em}\textcolor{brown}{nlayers}="100", \\
        \hspace*{1em}\textcolor{brown}{retrieval\_mode}="reduced", \\
        \hspace*{1em}\textcolor{brown}{output\_basename}="wasp39b/retrievals"\\
        )\\
        \textcolor{teal}{response:}
        \textcolor{gray}{TauREx Retrieval Complete!}\\

        \textcolor{gray}{Best-fit parameters:\\
          - planet\_radius: 1.2925015939305264\\
          - T: 1250.5205667982082\\
          - log\_H2O: -8.2292746957802\\
          - log\_CH4: -8.663445411550889\\
          - log\_CO: -7.854728018181991\\
          - log\_CO2: -8.197589024993974\\
          - log\_NH3: -6.9616671497038976}\\

        \textcolor{gray}{Log-likelihood: \{'local log-evidence': 649.9154831\}}\\

        \textcolor{gray}{Output files (in workspace):\\
          - fit\_png: wasp39b/retrievals/...\_multinest\_fit.png\\
          - corner\_png: wasp39b/retrievals/...\_multinest\_corner.png\\
          - wavelength\_npy: wasp39b/retrievals/...\_wavelength.npy\\
          - spectrum\_npy: wasp39b/retrievals/...\_spectrum.npy\\
          - samples\_npy: wasp39b/retrievals/...\_multinest\_samples.npy\\
          - weights\_npy: wasp39b/retrievals/...\_multinest\_weights.npy}
    \end{orchestraltoolmessage}

    \begin{orchestraltoolmessage}{DisplayImage}
        \textcolor{teal}{image:}\\
        wasp39b/retrievals/radius\_temp\_h2o\_ch4\_co\_co2\_nh3\_
        multinest\_fit.png (Figure~\ref{fig:fit_example})

    \end{orchestraltoolmessage}

    The atmospheric retrieval for WASP-39b using the multinest optimizer is complete. The retrieval fit parameters were the planet radius, temperature, and molecular abundance.\\

    Best-fit values obtained:\\

    \begin{itemize}
    \addtolength{\itemindent}{1cm}
      \item Planet radius: 1.293 Jupiter radii
      \item Temperature: 1251 K
      \item H2O log(abundance): approximately -8
      \item CH4 log(abundance): approximately -9
      \item CO log(abundance): approximately -8
      \item CO2 log(abundance): approximately -8
      \item NH3 log(abundance): approximately -7\\ 
    \end{itemize}

    I have displayed the fit plot showing the observed and retrieved spectrum. If you want, I can also show the corner plot for the posterior distributions or help with further analysis.

\end{orchestralagentmessage}

\section{Further results}
\label{sec:results}

In addition to the previous example interaction, in this section we showcase additional capabilities and present the complete set of results produced by the agent for the test case of WASP-39b. This includes figures generated throughout the analysis pipeline, shown in the order they were created, from the initial setup to the final retrieval outcomes. We follow the path of a typical exploratory process: extraction of planetary parameters, spectra binning experiment, visualization of observations and a comparison of two retrievals.\\

We begin by presenting the planetary and stellar parameters automatically extracted by the agent from NASA Exoplanet Archive for WASP-39b. These parameters define the physical setup of the system and serve as inputs for all subsequent modeling and retrieval steps. The quoted values were cross-checked and found to be consistent with the published reference parameters. They are shown in Table \ref{tab:system_parameters}.\\

\begin{table}
    \centering
    \begin{tabular}{l c c}
        \hline
        Parameter &  Value \\
        \hline
        Stellar mass            & 0.913 M$_\odot$       \\
        Stellar radius          & 0.939 R$_\odot$       \\
        Stellar effective temperature & 5485 K   \\
        \hline
        Planet mass             & 0.281 M$_{jup}$           \\
        Planet radius           & 1.279 R$_{jup}$          \\
        Planet equilibrium temperature &  1166 K     \\
        \hline
    \end{tabular}
    \caption{Planetary and stellar parameters for WASP-39b used in this study. Values were obtained automatically by the agent and verified against the NASA Exoplanet Archive.}
    \label{tab:system_parameters}
\end{table}

\begin{figure}
    \centering
    \includegraphics[width=\linewidth]{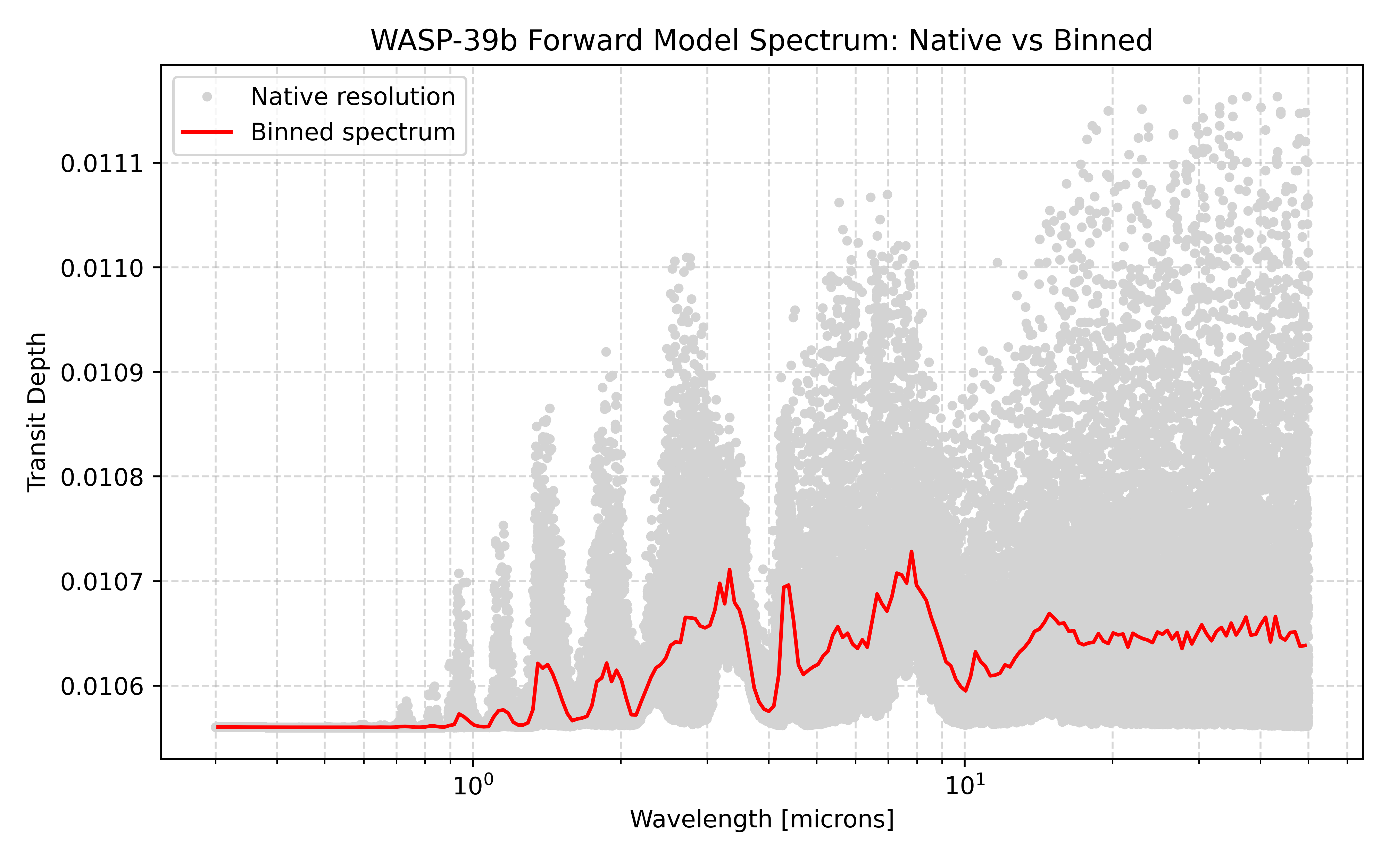}
    \caption{Transmission spectrum at native resolution in the background (gray points), overlaid with a binned spectrum (red curve) recalculated by the agent.
    \textbf{Prompt:} \emph{Can you bin down the spectrum and plot both resolutions on a same plot?}}
    \label{fig:native_and_binned_w39b}
\end{figure}

\begin{figure}
    \centering
    \includegraphics[width=\linewidth]{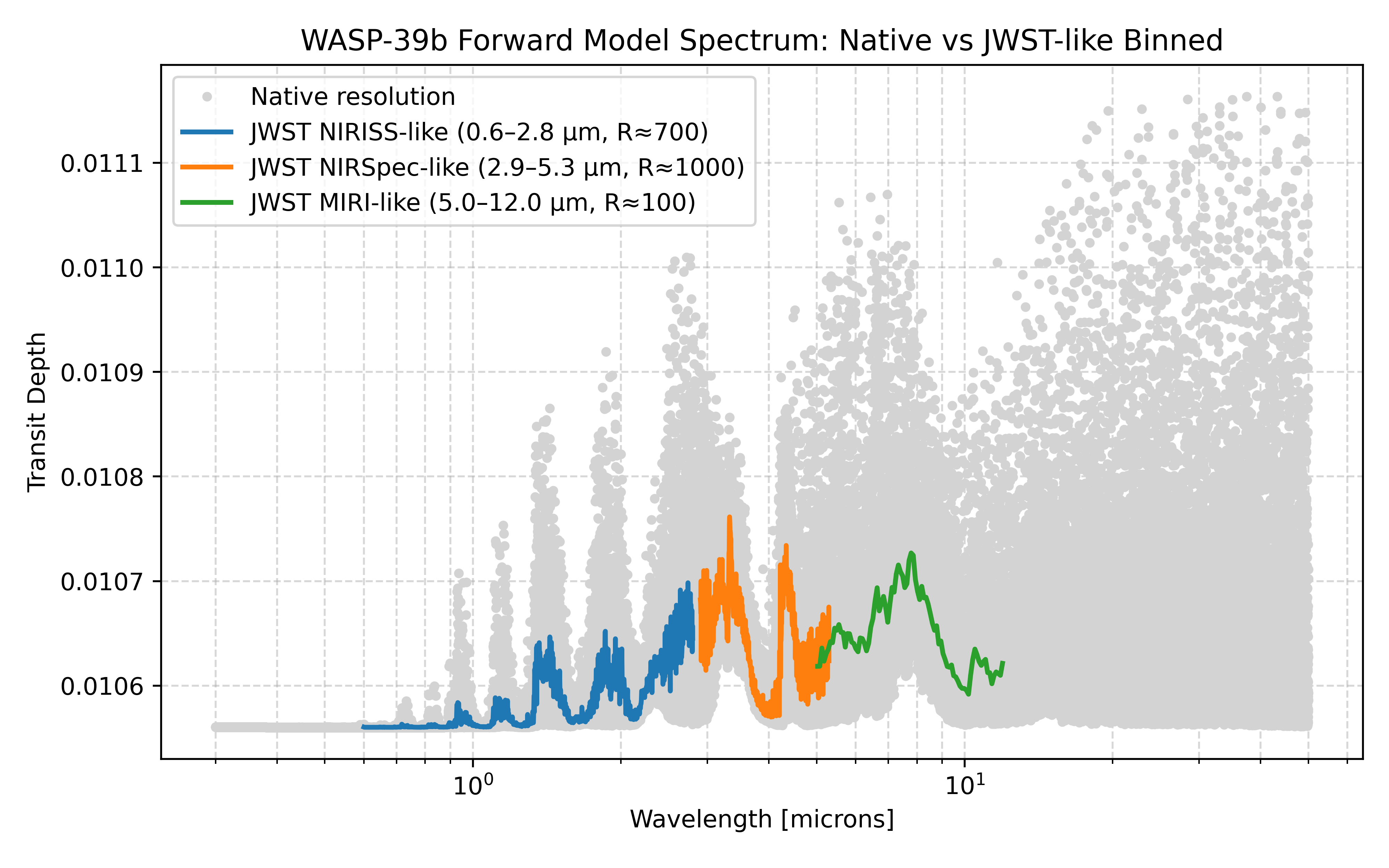}
    \caption{Transmission spectrum at native resolution in the background (gray points), overlaid with a binned spectrum recalculated by the agent. \newline 
    \textbf{Prompt:} \emph{Can you now bin the forward model down to JWST medium resolution?}}
    \label{fig:native_and_binned_jwst_w39b}
\end{figure}

\begin{figure}
    \centering
    \includegraphics[width=\linewidth]{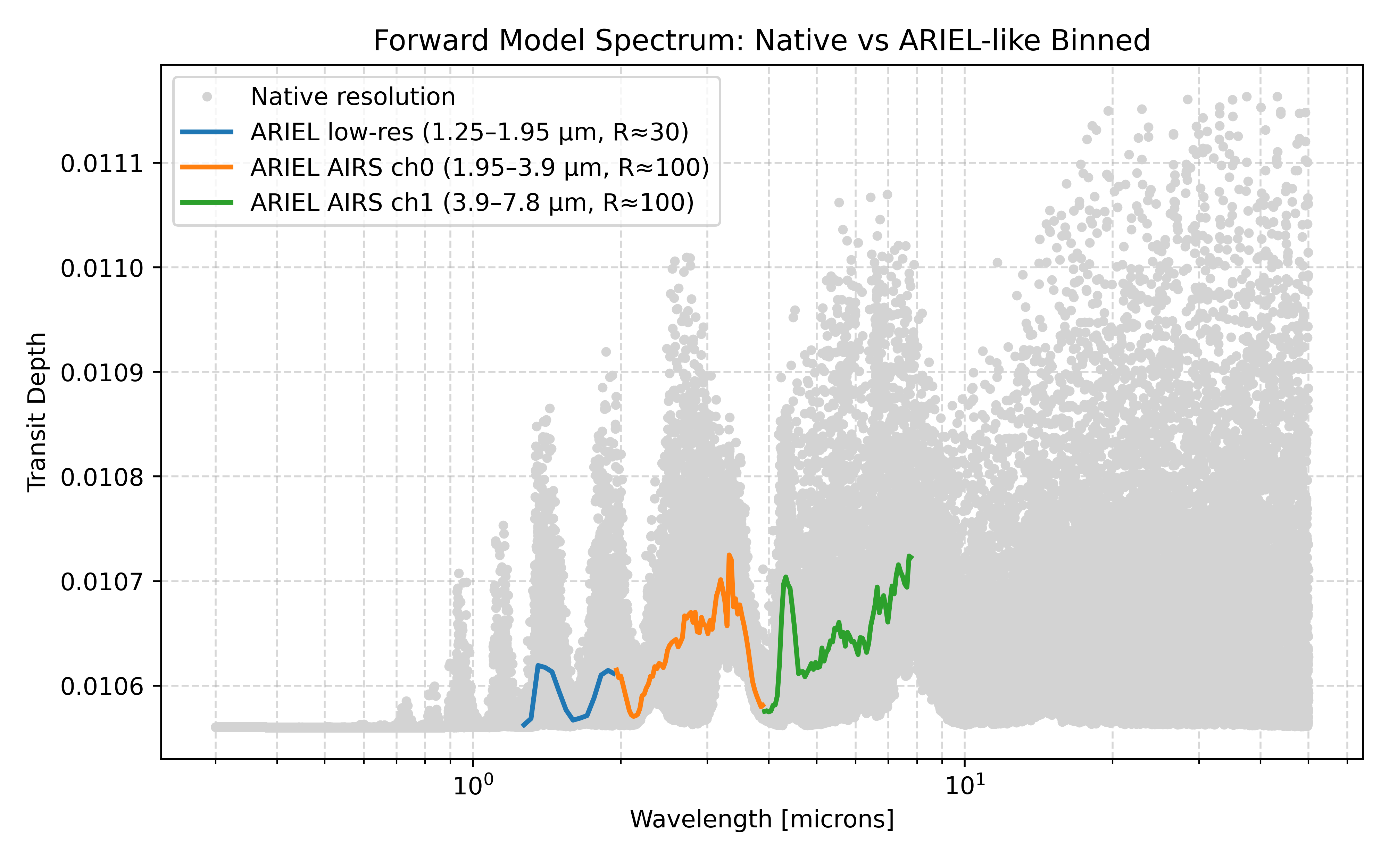}
    \caption{Transmission spectrum at native resolution in the background (gray points), with a binned down spectrum recalculated by the agent on top. 
    \newline
    \textbf{Prompt:} \emph{Can you now bin the forward model down to Ariel resolution?}}
    \label{fig:native_and_binned_ariel_w39b}
\end{figure}

We then tested the agent's ability to perform rebinning of the output spectrum generated by the \texttt{RunTaurexModel} tool. Note that ASTER currently does not have a dedicated rebinning tool and, the agent instead accomplished this task using only its built-in coding tools. This demonstrates its flexibility and ability to generalize from the available context. Using Table \ref{tab:system_parameters} system parameters, we prompted the agent to generate transmission spectra at multiple spectral resolutions, each prompt is specified within the caption of the respective figure (see Figures \ref{fig:native_and_binned_w39b}, \ref{fig:native_and_binned_jwst_w39b}, and \ref{fig:native_and_binned_ariel_w39b}). When no resolution is specified, the agent calculates the spectrum at the native model resolution, which is the resolution of the cross-section files (see Figure \ref{fig:display_example}). We requested the agent to reduce the resolution without specifying a value. The agent chose a resolution of R = 100, as shown in Figure \ref{fig:native_and_binned_w39b}. We also asked the agent to bin the native spectrum down to a JWST resolution and Ariel resolution. We emphasize that the agent was not given information about instruments design or spectral specifications (resolution or spectral range). The resulting spectra are shown in Figures \ref{fig:native_and_binned_jwst_w39b} and \ref{fig:native_and_binned_ariel_w39b}, which were directly generated by the agent. The agent decided to break the binned spectra into different JWST instruments and Ariel bands. For comparison, NIRISS SOSS covers approximately 0.6–2.8 $\mu$m with a wavelength dependent resolution ranging from R $\approx$ 700 at short wavelengths to R $\approx$ 1500 at longer wavelengths. The agent chose to approximate this mode using a representative resolution of R = 700 for this task. NIRSpec covers 0.6-5.3 $\mu$m with a resolution of R $\approx$ 100 in the prism mode, R $\approx$ 1000 in the medium resolution mode, and R $\approx$ 2700 in the high resolution mode. The agent chose to use the medium resolution mode (R $\approx$ 1000) for this task. MIRI covers approximately 5–12 $\mu$m in its low-resolution spectroscopic (LRS) mode with a resolution of R $\approx$ 100, the agent chose to use this resolution for the task. On the other hand, Ariel covers 1.25–7.8 $\mu$m using a combination of a low-resolution near-infrared spectrometer and the AIRS medium-resolution spectrometer. The near-IR spectrometer (1.25–1.95 $\mu$m) operates at low resolution with R $\approx$ 30, while the AIRS instrument is split into two channels covering 1.95–3.9 $\mu$m and 3.9–7.8 $\mu$m with a wavelength-dependent resolution of R $\approx$ 30–200. Accordingly, the agent chose to approximate these modes using representative resolutions for each band. This illustrates the abilities of the agent to perform common scientific tasks with minimal prompting, making use of the built-in Orchestral programming tools and leveraging the reasoning and planning capabilities of LLMs.\\

\begin{figure}
    \centering
    \includegraphics[width=\linewidth]{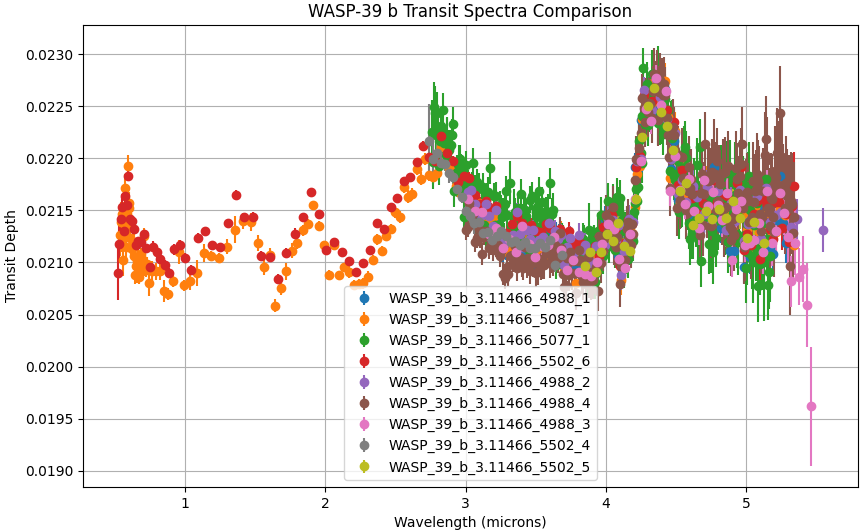}
    \caption{Comparison of all nine observations for WASP-39b with NIRSpec. These observations are public and are published in \protect\cite{JWST_ERS_2022,Alderson_2022,Rustamkulov_2023,Carter_2024}. This figure was produced by the agent. 
    \newline
    \textbf{Prompt:} \emph{Can you plot all the observations in a same figure for comparison?}}
    \label{fig:datasets_comparison_w39b}
\end{figure}

For the next exercise, we requested that the agent download observational datasets from the NASA Exoplanet Archive for WASP-39b, specifically using JWST NIRSpec instrument, and display them. 
Figure \ref{fig:datasets_comparison_w39b} presents the 9 datasets that were publicly available on NASA Exoplanet Archive \citep{JWST_ERS_2022,Alderson_2022,Rustamkulov_2023,Carter_2024}. The agent easily downloaded and plotted the selected observations.\\

\begin{table}
\centering
\begin{tabular}{lcc}
\hline
\textbf{Fitted Parameter} & \textbf{Retrieval 1} & \textbf{Retrieval 2} \\
\hline
$T$        & 1318.5 & 666.5 \\
$R_p$      & 1.29 & 1.30 \\
$\log \mathrm{H_2O}$ & -8.5 & -4.0 \\
$\log \mathrm{CH_4}$ & -8.8 & -7.1 \\
$\log \mathrm{CO}$   & -8.0 & -5.7 \\
$\log \mathrm{CO_2}$ & -8.7 & -5.0 \\
$\log \mathrm{NH_3}$ & -7.0 & -8.6 \\
\hline
\end{tabular}
\caption{Fitted parameters from the two retrieval runs. The results are discussed in Section \ref{sec:results}.}
\label{tab:retrievals}
\end{table}

\begin{figure}
    \centering
    \includegraphics[width=\linewidth]{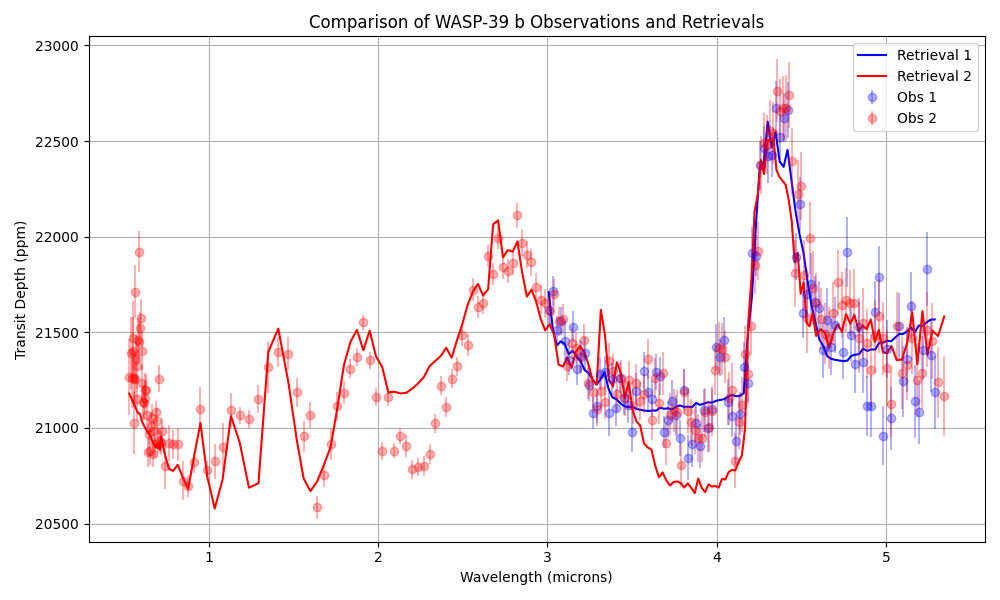}
    \caption{Comparison of the two retrieval results launched by the agent. The first observation used by the agent \protect\citep{JWST_ERS_2022} and the corresponding retrieved spectrum are shown in blue. The second observation used by the agent \protect\citep{Rustamkulov_2023} and the corresponding retrieved spectrum are shown in red. 
    \newline
    \textbf{Prompt:} \emph{Can you do a plot with the two observations and the two retrieved spectra so I can compare the results?}}
    \label{fig:retrieval_results_fits}
\end{figure}

\begin{figure}
    \centering
    \includegraphics[width=\linewidth]{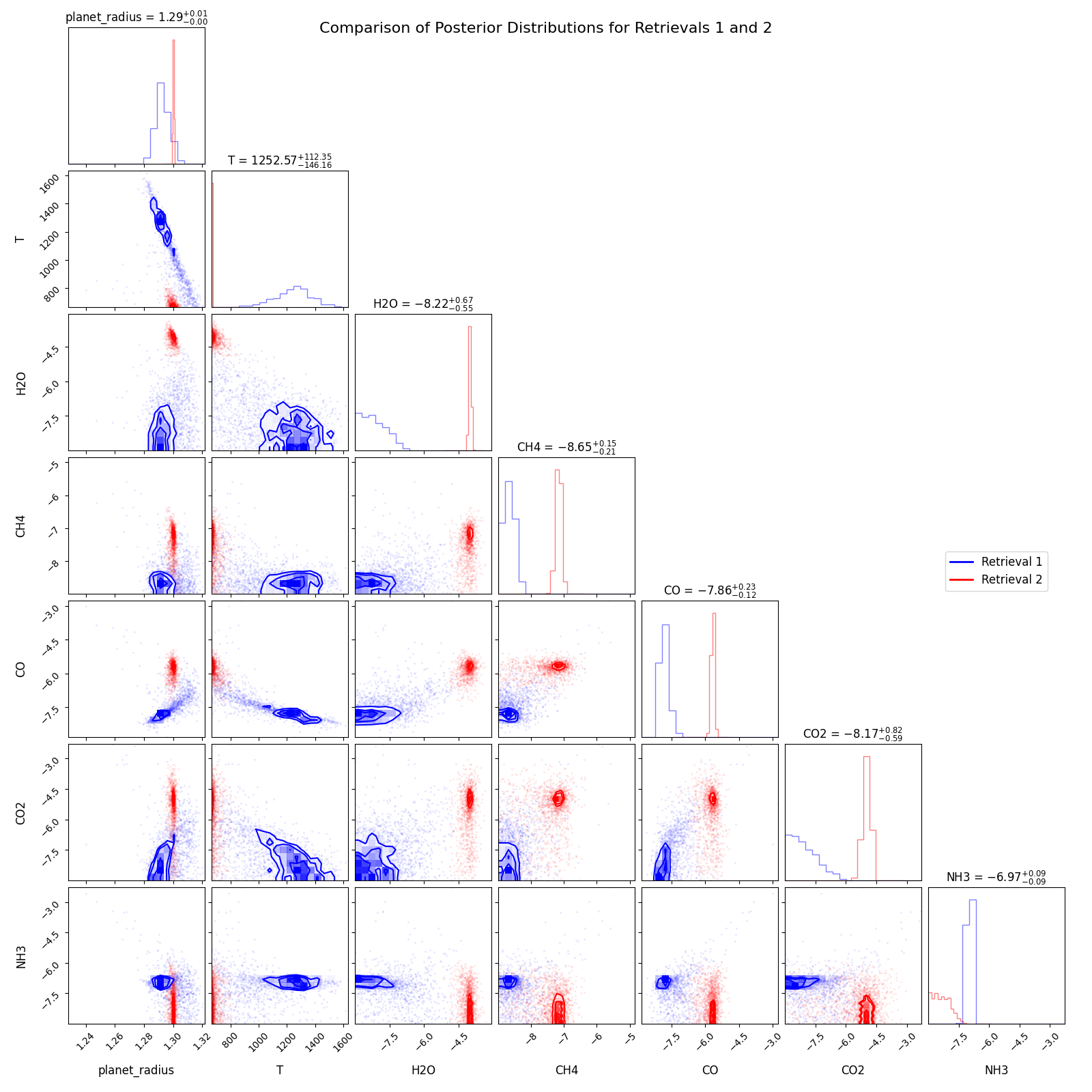}
    \caption{Corner plot of the posterior distributions of the fitted parameters from the retrievals launched by the agent. The blue distributions are corresponding to the first retrieval on \protect\cite{JWST_ERS_2022} data. The red distributions are corresponding to the second retrieval on \protect\cite{Rustamkulov_2023} data. 
    \newline
    \textbf{Prompt:} \emph{Can you now do a corner plot with the probability distribution results of the two retrievals on the same corner plot figure?} }
    \label{fig:retrieval_results_corner}
\end{figure}

We next asked the agent to perform a retrieval analysis on two of the JWST observations. For the first retrieval, the agent used an observation published by \cite{JWST_ERS_2022} and for the second retrieval the observation by \cite{Rustamkulov_2023}. The resulting best-fit models together with the observations and the posterior distributions for the fitted parameters are shown in Figures \ref{fig:retrieval_results_fits} and \ref{fig:retrieval_results_corner}. The agent was given instructions to retrieve the planet temperature, planet radius, H$_2$O, CH$_4$, CO, CO$_2$, and NH$_3$ abundances. The bounds were not explicitly specified but some recommendations are provided in the \texttt{retrieval\_best\_practices} skill, which captures the domain expert prior experience (see Section \ref{sec:skills}). The recommended planet radius bounds were $\pm$0.5 of the radius value, and the recommended temperature bounds were $\pm$500 of the temperature value. The molecular abundance bounds were set to [$10^{-9}$;$10^{-2}$], following the skill recommendation. The agent used the Multinest optimizer option (recommended in the skill file) with the minimum number of live points set to 100. For this proof of concept exercise, we limited the chemical inventory to the listed molecules above. The agent proceeded to perform both retrievals autonomously without further prompting. For Figure \ref{fig:retrieval_results_corner} the agent followed the plotting recommendations described in the \texttt{corner\_plots} skill, and displayed both retrieval results on the same figure.
Figure \ref{fig:retrieval_results_corner} shows the posterior distributions for each fitted parameter for both retrievals. While the focus of the present study is to demonstrate the technical capabilities of an agentic workflow, and an in-depth analysis of the observations is beyond the scope of this paper, one can see that both retrievals favor different solutions. The red retrieval generally exhibits narrower, more peaked distributions, meaning that those parameters are better constrained with smaller uncertainties, whereas the blue retrieval has broader distributions, indicating weaker constraints and stronger degeneracies. This is expected since the \cite{JWST_ERS_2022} data spans 3.0-5.5 $\mu$m, while \cite{Rustamkulov_2023} data spans 0.5-5.5 $\mu$m with NIRSpec's PRISM mode. We can also see that the red and blue posterior distributions are offset, indicating possible disagreements. This mismatch is also expected because wider wavelength coverage includes additional spectral features and probes different atmospheric layers, shifting the best-fit values. The planet radius remains the most consistent between the two retrievals because it primarily controls the overall transit depth normalization, which is well constrained even with limited wavelength coverage and does not strongly depend on detailed spectral features. One can improve the quality of the fit by adding a more comprehensive molecular list in the retrieval model, consistent with previous studies. 

In this work, our objective was to demonstrate the feasibility and efficiency of using an agent to perform end-to-end retrieval analyses. In that respect, the results presented here successfully validate the capabilities of ASTER: the agent is able to autonomously handle data acquisition, model generation, retrieval execution, and result interpretation within a realistic scientific workflow. This proof of concept application to WASP-39b demonstrates that agents can be effectively integrated with existing exoplanet tools.

\section{Conclusion}
\label{sec:conclusion}    

In this paper, we have presented the Agentic Science Toolkit for Exoplanet Research. The goal of ASTER is not to replace existing frameworks, but to act as an assistant that can coordinate common tasks and provide suggestions. In this work, we applied ASTER to data acquisition, modeling, retrieval, and visualization in the context of transmission spectroscopy of exoplanets' atmospheres. 

It is important to note that the goal of the present study is not to improve upon previous analyses of WASP-39b's NIRSpec observations. These observations have been extensively studied by \cite{royperez2026wasp39b,Roy_Perez_2025,esparzaborges2025wasp39b,esparzaborges2023wasp39b,Grant_2023,Rustamkulov_2023,Alderson_2022}. Instead, our goal is to illustrate the capabilities of ASTER on a realistic test case and demonstrate that an agent can effectively conduct core tasks in exoplanet research workflows. We have shown that the agent can autonomously orchestrate the full pipeline, from obtaining planetary parameters and downloading datasets from public repositories to generating transmission spectra from radiative transfer models, performing Bayesian retrievals of atmospheric parameters, and visualizing the results. The successful autonomous execution of multiple retrievals using different datasets, along with their comparison, highlights the flexibility and utility of the approach. In this sense, the goal of the study is achieved: we demonstrate that an agent can reliably integrate existing exoplanet tools into a coherent, reproducible, and efficient analysis framework.

Unlike a fixed script, which encodes a predefined sequence of operations, ASTER dynamically adapts its behavior based on intermediate results, user input, and evolving context, enabling more flexible and interactive scientific workflows. Additionally, the LLM retains a broad base of knowledge from its training data, which can be useful when a query is under-specified or when a tool call implicitly relies on prior knowledge. For example, ASTER was not explicitly provided with the spectral resolution of Ariel, which was required for the rebinning experiment; nevertheless, it was able to infer this information when needed. However, practitioners should not rely solely on this implicit knowledge, providing explicit context remains best practice. In situations where the model must infer missing details, the risk of hallucination increases. 

Another key strength of the agent-based approach is format compatibility. Oftentimes, data comes in different formats and units, and making different inputs/outputs compatible is tedious. By standardizing how datasets, models, and retrieval outputs are ingested and produced, ASTER simplifies comparisons between different analyses and results. For example, comparisons between different retrieval models are highly non-trivial, as they often differ in implementation choices, statistical methods, evaluation metrics, or stopping criteria \citep{Barstow_2020_comp_retrievals,Mugnai_2024}.

ASTER is also modular by design. Individual capabilities are  independent tools, allowing new functionality to be added without requiring modification of the overall system architecture. This makes ASTER adaptable to evolving scientific needs, such as new instruments, updated chemical models, or alternative retrieval techniques. 

Beyond orchestration, ASTER functions as a research assistant. The agent actively checks intermediate results, validates inputs, and verifies assumptions when possible. When errors occur, the agent attempts to diagnose issues, suggest corrective actions, and retry failed steps. In addition, the agent can highlight unexpected or atypical behavior in results, prompting further inspection by the user. Routine and repetitive tasks, such as parameter explorations, resolution changes, or repeated retrievals, can be automated, allowing researchers to focus on interpretation rather than execution.

Another important aspect of ASTER is its accessibility. By lowering the technical barrier to entry, this toolkit makes advanced analyses such as atmospheric retrievals  more approachable for early-career researchers and non-specialists. 

Finally, ASTER represents a first step toward more capable scientific agents. The framework demonstrates how agent-based systems can support rigorous scientific analysis. As tools, models, and datasets continue to grow in complexity, agent-based systems may play an increasingly important role in managing workflows, enforcing best practices, and enabling more efficient and reproducible research. We strongly encourage the exoplanet community to extend this library of tools and implement new capabilities  within the ASTER framework.

\section*{Acknowledgments}

The work of EP, AR and KTM is supported in part by the Shelby Endowment for Distinguished Faculty at the University of Alabama. The authors thank Tony Menzo for insightful discussions on agents and the development of ASTER.
LP acknowledges support from the scholarship co-financed by the Ministerial Decree no. 118 of 2nd March 2023,760 based on the NRRP - funded by the European Union - NextGenerationEU - Mission 4 Component 1 – CUP C96E23000340001.

\bibliographystyle{rasti}
\bibliography{references}

\end{document}